\documentclass{aa}
\usepackage[utf8]{inputenc}
\usepackage{natbib}
\usepackage{graphicx}
\usepackage{amsmath}
\usepackage{gensymb}
\usepackage{subcaption}
\usepackage{dblfloatfix}
\usepackage{makecell}
\usepackage{ragged2e}
\usepackage[colorlinks=true,allcolors=blue]{hyperref}
\usepackage[scientific-notation=true]{siunitx}
\usepackage{xcolor}

\title{The mass and size of Herbig disks as seen by ALMA}

\author{L. M. Stapper\inst{\ref{inst1}}\fnmsep\thanks{E-mail: \texttt{stapper@strw.leidenuniv.nl}} \and M. R. Hogerheijde\inst{\ref{inst1}, \ref{inst2}} \and E. F. van Dishoeck\inst{\ref{inst1},\ref{inst3}} \and R. Mentel\inst{\ref{inst1}, \ref{inst4}}}

\institute{Leiden Observatory, Leiden University, PO Box 9513, 2300 RA Leiden, The Netherlands \label{inst1} \and Anton Pannekoek Institute for Astronomy, University of Amsterdam, PO Box 94249, 1090 GE, Amsterdam, The Netherlands \label{inst2} \and Max-Planck-Institut für Extraterrestrische Physik, Giessenbachstrasse 1, 85748 Garching, Germany \label{inst3} \and School of Physics, University College Dublin, Belfield, Dublin 4, Ireland \label{inst4}}

\date{\today}

\abstract
{ 
Many population studies have been performed over the past decade with the Atacama Large millimeter/submillimeter Array (ALMA) to understand the bulk properties of protoplanetary disks around young stars. The studied populations have mostly consisted of late spectral type (i.e., G, K \& M) stars, with relatively few more massive Herbig stars (spectral types B, A \& F). With GAIA-updated distances, now is a good time to use ALMA archival data for a Herbig disk population study and take an important step forward in our understanding of planet formation.
}
{ 
The aim of this  work is to determine the masses and sizes of all Herbig dust disks observed with ALMA to date in a volume-limited sample out to 450~pc. These masses and sizes are put in the context of the Lupus and Upper~Sco T~Tauri disk populations.
}
{ 
ALMA Band 6 and Band 7 archival data of 36 Herbig stars are used, making this work $64\%$ complete out to 225~pc, and 38\% complete out to 450~pc also including Orion. Using stellar parameters and distances, the dust masses and sizes of the disks are determined via a curve-of-growth method. Survival analysis is used to obtain cumulative distributions of the dust masses and radii.
}
{ 
Herbig disks have a higher dust mass than the T~Tauri disk populations of Lupus and Upper~Sco by factors of $\sim3$ and $\sim7$ respectively. In addition, Herbig disks are often larger than the typical T~Tauri disk. Although the masses and sizes of Herbig disks extend over a similar range to those of T~Tauri disks, the distributions of masses and sizes of Herbig disks are significantly skewed toward higher values. Lastly, group I disks are more massive than group II disks. An insufficient number of group II disks are observed at sufficient angular resolution to determine whether or not they are also small in size compared to group I disks.
}
{ 
Herbig disks are skewed towards more massive and larger dust disks compared to T~Tauri disks. Based on this we speculate that these differences find their origin in an initial disk mass that scales with the stellar mass, and that subsequent disk evolution enlarges the observable differences, especially if (sub)millimeter continuum optical depth plays a role. Moreover, the larger disk masses and sizes of Herbig stars could be linked to the increasing prevalence of giant planets with host star mass.
}


\begin{document}

\maketitle

\section{Introduction}
\label{sec:introduction}

\begin{figure*}[h!]
    \centering
    \includegraphics[width=0.98\textwidth]{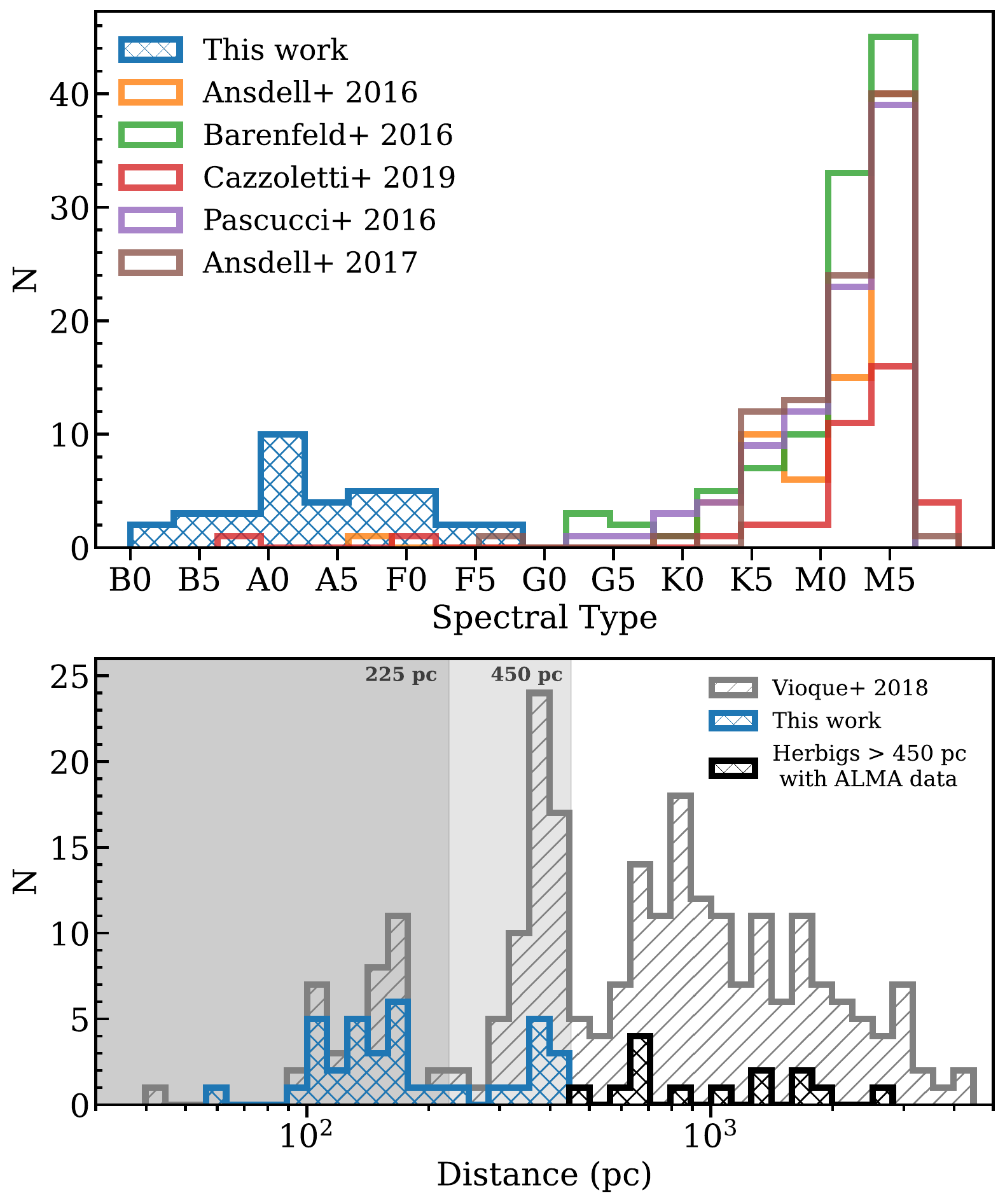}
    \caption{\textbf{Top:} Histogram of spectral types of five different population studies together with the full sample of this work. The population studies included are: Lupus \citep{Ansdell2016}, $\sigma$ Orionis \citep{Ansdell2017}, Upper~Sco \citep{Barenfeld2016}, Corona Australis \citep{Cazzoletti2019}, and Chamaeleon I \citep{Pascucci2016}. \textbf{Bottom:} Histogram of the distances of the 252 Herbig disks as defined by \citet{Vioque2018} and the full sample used in this work. The black histogram shows the Herbig disks beyond 450~pc for which ALMA data are available. The shaded areas indicate the two different cutoffs used in this work.}
    \label{fig:Herbig_sample_hist}
\end{figure*}

Many population studies of protoplanetary disks in various star-forming regions have been carried out in recent years with the Atacama Large millimeter/submillimeter Array (ALMA). These demographic studies have led to new insights into disk evolution by finding relations between stellar mass and disk mass \citep[e.g.,][]{Andrews2013, Barenfeld2016, Ansdell2016, Pascucci2016}, and between disk radius and disk mass \citep{Andrews2013, Andrews2018a, Tripathi2017, Tazzari2021}. On the other side of the planet formation process, exoplanet demographics show a trend where higher mass stars (1~M$_\odot$ and higher) are more often associated with planets more massive than $100$~M$_\oplus$ than stars with lower stellar masses \citep[e.g.,][]{Johnson2007, Johnson2010, Wittenmyer2020, Fulton2021}. The increasing number of observed exoplanets and disks allows for connections to be made between the beginning and ending stages of planet formation \citep[e.g][]{Mulders2021}, improving our understanding of the relations found in both disk and exoplanet demographics.


The increase in occurrence rate of giant exoplanets around more massive main sequence stars coincides with the frequency of substructure in disks ---as seen in the (sub)millimeter continuum emission--- in the form of rings and gaps \citep{vanderMarel2021}, suggesting a connection between the two. One pathway for these structures to form is via planets forming inside these disks, creating pressure bumps preventing the radial drift of dust \citep[e.g.,][]{Birnstiel2010} and influencing the overall evolution of a disk \citep{Pinilla2020, Cieza2021}. The relationships found between stellar mass, disk mass, and disk extent might explain the observed increase in occurrence rate of massive planets with stellar mass and orbital radius \citep{Andrews2013}. This introduces an enticing scenario in which the more massive disks produce gas giant planets at greater distances, creating structured disks, while the less massive disks shrink over time because of radial drift \citep{Cieza2021, vanderMarel2021}. Indeed, the more massive disks seem to retain their disk mass for longer, steepening the stellar mass--dust mass relation over time \citep{Pascucci2016, Ansdell2017}, while on the contrary, the dust radii in the 5--11~Myr-old Upper~Sco star-forming region (SFR) were found to be significantly smaller than those in the younger 1--3~Myr-old Lupus region \citep{Barenfeld2017, Ansdell2018}, suggesting an evolution towards smaller disks.

However, as expected from the initial mass function \citep[e.g.,][]{Kroupa2001, Bastian2010}, these disk population studies mainly consist of T~Tauri stars and lack coverage of more massive pre-main sequence (PMS) stars (cf. top panel of Fig. \ref{fig:Herbig_sample_hist} and Table \ref{tab:population_studies}). These more massive stars, known as Herbig Ae/Be stars \citep{Herbig1960, Waters1998}, typically have masses between $2$~M$_\odot$ and $12$~M$_\odot$, are {optically} visible and have, by definition, an excess in infrared (IR) light because of a disk surrounding them. Infrared excess is used to classify their spectral energy distributions (SED) into group I and group II disks \citep{Meeus2001}. This division was based on whether the SED could be described by a single power law at mid- to far-IR wavelengths or if an additional blackbody component was necessary. Lastly, due to the PMS evolutionary tracks being horizontal for early-spectral-type stars, stars which end up with a spectral type A will first be F- or even G-type stars. Therefore, Herbig Ae stars are relatively old. To counteract this, research into intermediate-mass T~Tauri stars ---which precede Herbig Ae stars--- has been done \citep[recently e.g.,][]{Valegard2021} and the definition of Herbig Ae/Be stars is often expanded to include F-type stars \citep[e.g.,][]{Chen2016, Vioque2018}.

While T~Tauri stars are abundant throughout star-forming regions, less than 300 Herbig stars are known to date \citep{The1994, Carmona2010, Chen2016, Vioque2018, Vioque2020, GuzmanDiaz2021}, which include stars with spectral types ranging from B to F. Before ALMA, dust mass studies of disks around Herbig stars (hereafter: Herbig disks) have been done with other facilities such as the NRAO Very Large Array (VLA) and the IRAM Plateau de Bure \citep[e.g.,][]{AlonsoAlbi2009} and while recent work by \citet{vanderMarel2021} does include all Herbig disks associated with nearby SFRs (but does not include isolated Herbig disks), there has not yet been a dedicated population study of Herbig disks observed with ALMA. Nevertheless, a significant portion of the nearby Herbig disks have been observed with ALMA over the years in a variety of programs and these data are used in the present work for a first systematic look at Herbig disks with ALMA.

In this study, we use a more complete sample of Herbig disks compared to any other survey done before. Distances from GAIA parallaxes and newly determined stellar properties \citep{Vioque2018} allow for a well-defined sample. What is the typical extent and mass of a Herbig disk? Do the dust disk mass and extent follow the expected stellar mass--dust mass relationship? Do Herbig disks fall within the posed scenario in which larger disks form giant planets further out, while smaller disks decrease in size over time? Are there any differences between the dust mass and extent and the groups defined by \citet{Meeus2001} based on their SEDs? To answer these questions, we determine the dust mass and extent of Herbig disks with ALMA archival data in a volume-limited sample and compare these parameters with their T~Tauri counterparts from previous population studies. Section \ref{sec:sample_selection} shows the selection process of the sample used in this work. Section \ref{sec:data_and_reduction} sets out how the data are retrieved and how the integrated (sub)millimeter flux and size was determined. Section \ref{sec:results} describes the results of the dust mass and extent measurements and compares these to the Lupus and Upper~Sco SFRs. In Sect. \ref{sec:discussion} we discuss the implications on the stellar mass--dust mass and disk radius--dust mass relations and look at the \citet{Meeus2001} group dichotomy. Section \ref{sec:conclusion} summarizes our conclusions.

\begin{figure*}
    \centering
    \includegraphics[width=0.9\textwidth]{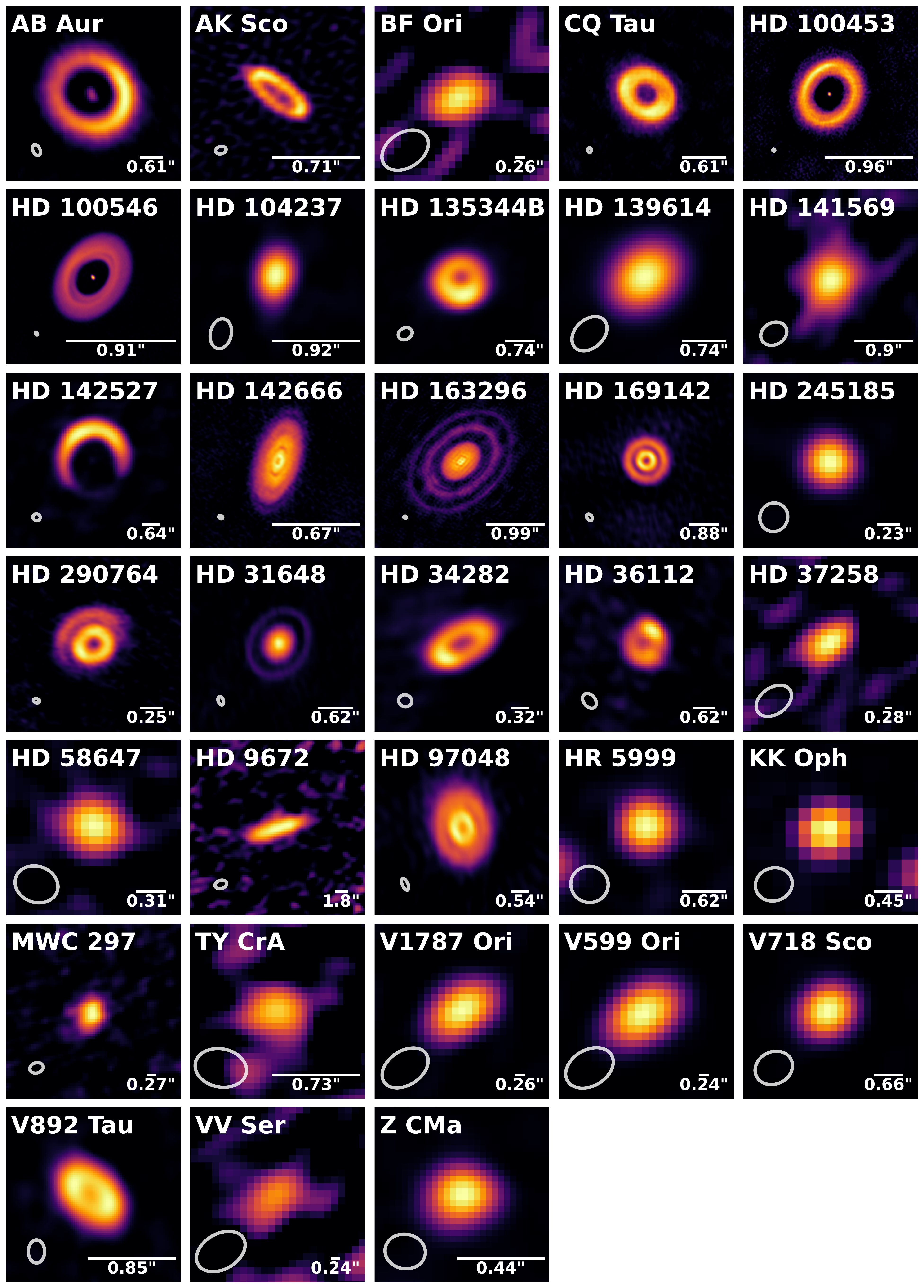}
    \caption{ALMA Band 6 and 7 continuum images of all objects. The size of the beam is shown in the bottom left of each plot and a scale bar of 100~au in size together with the angular scale in arcseconds is shown in the bottom right. Each image is normalized with an asinh stretch to make the fainter details of the disk more visible. HD~53367, HD~176386 and R~CrA, which are not detected, are not shown.}
    \label{fig:gallery}
\end{figure*}

\begin{table*}[t]
\caption{Coordinates and spectral types of the Herbig stars used in this work and the calculated Band 6 and 7 flux densities and dust masses of each Herbig disk.}
\tiny
\begin{tabular}{l l r l l c c c c c c l}
\hline\hline
\makecell{Name \\ \hspace{1mm}} & \makecell{R.A.$_\text{J2000}$ \\ (h:m:s)} & \makecell{Decl.$_\text{J2000}$ \\ (deg:m:s)} & \makecell{Sp.Tp. \\ \hspace{1mm}} & \makecell{Group \\ \hspace{1mm}} & \makecell{$F_\text{cont.}$ \\ (mJy)} & \makecell{$M_\text{dust}$ \\ ($M_\oplus$)} & \makecell{$R_\text{dust, 68\%}$ \\ (au)} & \makecell{$R_\text{dust, 90\%}$ \\ (au)} & \makecell{inc \\ ($\degree$)} & \makecell{PA \\ ($\degree$)} & \makecell{Ref. \\ \hspace{1mm}} \\ \hline
AB Aur$^\alpha$            & 04:55:45.9 & +30:33:04 & A0   & I  & 192.8  & 11.8±1.2   & 186±15 & 225±34  & 23.2 & 99    & 1, i, $\beta$      \\
AK Sco                     & 16:54:44.9 & -36:53:19 & F5   & II & 26.1   &  6.1±0.6   & 42±3   & 53±15   & 71   & 39    & 2, ii, $\beta$     \\
BF Ori                     & 05:37:13.3 & -06:35:01 & A7   & II & 0.8    &  1.1±0.1   & <233   & <311    & --   & --    & 3, $\beta$         \\
CQ Tau$^\alpha$            & 05:35:58.5 & +24:44:54 & F5   & I  & 144.0  & 44.2±4.8   & 62±4   & 76±18   & 35   & 35    & 4, iii, $\beta$    \\
HD 100453                  & 11:33:05.6 & -54:19:29 & A9   & I  & 141.1  & 17.5±1.8   & 39±2   & 44±6    & 38   & 128   & 5, iv, $\beta$     \\
HD 100546                  & 11:33:25.4 & -70:11:41 & A0   & I  & 399.6  & 38.0±3.9   & 35±2   & 41±4    & 43   & 131   & 6, v, $\beta$     \\
HD 104237$^\alpha$         & 12:00:05.1 & -78:11:35 & A0   & II & 88.9   & 10.5±1.1   & <59    & <98     & --   & --    & 5, $\beta$         \\
HD 135344B                 & 15:15:48.5 & -37:09:16 & F8   & I  & 539.7  & 35.2±3.8   & 81±10  & 109±31  & 11   & 28    & 7,vi, $\beta$    \\
HD 139614                  & 15:40:46.4 & -42:29:54 & A9   & I  & 197.2  & 41.7±4.3   & <65    & <97     & --   & --    & 6, $\beta$        \\
HD 141569                  & 15:49:57.8 & -03:55:16 & A2   & II & 3.4    &  0.36±0.04 & <139   & <232    & --   & --    & 6, $\beta$        \\
HD 142527                  & 15:56:41.9 & -42:19:23 & F6   & I  & 1048.1 & 214.9±22.1 & 208±9  & 236±33  & 27   & 71    & 8, vii, $\beta$  \\
HD 142666                  & 15:56:40.0 & -22:01:40 & F0   & II & 118.6  & 25.1±2.6   & 40±3   & 52±11   & 60   & 109   & 6, viii, $\beta$   \\
HD 163296                  & 17:56:21.3 & -21:57:22 & A1   & II & 541.3  & 46.7±5.0   & 71±6   & 101±16  & 45   & 137   & 4, ix, $\beta$     \\
HD 169142$^\alpha$         & 18:24:29.8 & -29:46:49 & F1   & I  & 203.7  & 22.9±2.4   & 71±7   & 140±77  & 13   & 5     & 6, x, $\beta$   \\
HD 176386$^\alpha$         & 19:01:38.9 & -36:53:27 & B9   & II & <0.32  & <0.06      & --     & --      & --   & --    & 9, $\gamma$       \\
HD 245185                  & 05:35:09.6 & +10:01:52 & A0   & I  & 34.6   & 41.5±7.6   & <76    & <126    & --   & --    & 10, $\beta$        \\
HD 290764                  & 05:38:05.3 & -01:15:22 & F0   & I  & 210.1  & 90.3±11.8  & 119±16 & 167±28  & 30   & 154.6 & 5, xi, $\beta$   \\
HD 31648$^\alpha$          & 04:58:46.3 & +29:50:37 & A5   & II & 253.0  & 70.9±7.7   & 58±18  & 107±49  & 37   & 122   & 4, xii, $\beta$    \\
HD 34282                   & 05:16:00.5 & -09:48:35 & B9.5 & I  & 99.0   & 86.8±9.7   & 189±18 & 260±77  & 60   & 150   & 11, xiii, $\beta$     \\
HD 36112$^\alpha$          & 05:30:27.5 & +25:19:57 & A8   & I  & 70.7   & 18.8±2.0   & 90±17  & 123±50  & 21   & 65    & 5, xiv, $\beta$    \\
HD 37258$^\alpha$          & 05:36:59.3 & -06:09:16 & A3   & II & 1.8    &  2.4±0.4   & <326   & <435    & --   & --    & 3, $\beta$         \\
\textit{HD 53367}$^\alpha$ & 07:04:25.5 & -10:27:16 & B0   & ?  & <0.81  & <0.05      & --     & --      & --   & --    & 12                 \\
HD 58647$^\alpha$          & 07:25:56.1 & -14:10:44 & B9   & II & 1.9    &  1.0±0.1   & <75    & <126    & --   & --    & 13, $\beta$         \\
HD 9672$^\alpha$           & 01:34:37.9 & -15:40:35 & A1   & II & 4.1    &  0.13±0.01 & 315±39 & 473±111 & 81   & 161   & 13, xv, $\beta$     \\
HD 97048                   & 11:08:03.3 & -77:39:18 & A0   & I  & 711.4  & 155.9±16.0 & 182±19 & 245±69  & 41   & 85    & 14, xvi, $\beta$   \\
HR 5999                    & 16:08:34.3 & -39:06:18 & A7   & II & 26.5   &  4.0±0.4   & <48    & <81     & --   & --    & 15, $\beta$        \\
KK Oph$^\alpha$            & 17:10:08.1 & -27:15:19 & A6   & II & 27.7   & 17.4±2.5   & <88    & <88     & --   & --    & 16, $\delta$       \\
MWC 297                    & 18:27:39.5 & -03:49:52 & B1.5 & I  & 288.7  & 65.7±9.6   & <169   & <253    & --   & --    & 17, $\beta$        \\
\textit{R CrA}$^\alpha$    & 19:01:53.7 & -36:57:09 & B5   & II & <65.8  & <13.2 & --   & --    & --   & --    & 18, $\epsilon$     \\
TY CrA$^\alpha$            & 19:01:40.8 & -36:52:34 & B9   & I  & 0.67   & 0.10±0.01  & <27    & <34     & --   & --    & 5, $\beta$         \\
V1787 Ori$^\alpha$         & 05:38:09.3 & -06:49:17 & A5   & I  & 14.8   & 24.2±2.9   & <313   & <391    & --   & --    & 5, $\beta$         \\
V599 Ori                   & 05:38:58.6 & -07:16:46 & A8.9 & I  & 55.9   & 75.0±8.6   & <328   & <410    & --   & --    & 19, $\beta$        \\
V718 Sco                   & 16:13:11.6 & -22:29:07 & A5   & II & 49.5   & 11.9±1.3   & <69    & <114    & --   & --    & 20, $\beta$        \\
\textit{V892 Tau}$^\alpha$ & 04:18:40.6 & +28:19:15 & A0   & I  & 285.4  & 76.1±8.3   & 40±4   & 52±13   & 54.5 & 52.1  & 21, xvii, $\zeta$ \\
VV Ser                     & 18:28:47.9 & +00:08:40 & B5   & II & 2.3    & 2.3±0.3    & <239   & <399    & --   & --    & 22, $\beta$        \\
\textit{Z CMa}$^\alpha$    & 07:03:43.2 & -11:33:06 & B5   & I  & 29.3   & 7.6±6.4    & <28    & <41     & --   & --    & 23, $\beta$        \\ \hline
\end{tabular}\\
\textbf{Notes:} All distances are obtained from \citet{Vioque2018}. The luminosities, stellar masses, and ages were obtained from \citet{Wichittanakom2020} except the objects marked with an $\alpha$ of which these parameters were obtained from \citet{Vioque2018}. The objects in italics are from the \citet{Vioque2018} low-quality parallax sample, and the others are from the high-quality sample. The error on the continuum flux is $10\%$ of the given value. \textit{Spectral type references:} (1) \citet{Mooley2013}, (2) \citet{Houk1982}, (3) \citet{Tisserand2013}, (4) \citet{Mora2001}, (5) \citet{Vieira2003}, (6) \citet{Gray2017}, (7) \citet{Coulson1995}, (8) \citet{Houk1978}, (9) \citet{Torres2006}, (10) \citet{Gray1998}, (11) \citet{Houk1999}, (12) \citet{Tjin2001}, (13) \citet{Houk1988}, (14) \citet{Irvine1977}, (15) \citet{Bessell1972}, (16) \citet{Carmona2007}, (17) \citet{Drew1997}, (18) \citet{Gray2006}, (19) \citet{Hsu2013}, (20) \citet{Carmona2010}, (21) \citet{Skiff2014}, (22) \citet{GuzmanDiaz2021}, (23) \citet{Covino1984}. \textit{Inclination and position angle references:} (i) \citet{Tang2017}, (ii) \citet{Czekala2015}, (iii) \citet{UbeiraGabellini2019}, (iv) \citet{Benisty2017}, (v) \citet{Pineda2019}, (vi) \citet{Stolker2016}, (vii) \citet{Kataoka2016}, (viii) \citet{Rubinstein2018}, (ix) \citet{deGregorioMonsalvo2013}, (x) \citet{Fedele2017}, (xi) \citet{Kraus2017}, (xii) \citet{Liu2019}, (xiii) \citet{vanderPlas2017a}, (xiv) \citet{Isella2010}, (xv) \citet{Hughes2017}, (xvi) \citet{vanderPlas2017b}, (xvii) \citet{Long2021}. \textit{Meeus group references:} ($\beta$) \citet{GuzmanDiaz2021}, ($\gamma$) \citet{Boersma2009}, ($\delta$) \citet{Juhasz2010}, ($\epsilon$) \citet{Acke2004}, ($\zeta$) \citet{Menu2015}.
\label{tab:Herbig_params}
\end{table*}


\section{Sample selection}
\label{sec:sample_selection}
The sample in this work is based on the GAIA DR2 study of Herbig Ae/Be stars of \citet{Vioque2018} for which ALMA data are available. Previous population studies were limited to one SFR and consequently all stars have approximately the same distance, but the Herbig stars are distributed over the sky and therefore each has a different distance, which should be well determined. The newly determined GAIA distances and accurately derived stellar properties inspired us to do a population study on these objects.

\citet{Vioque2018} obtained a sample of 218 objects classified as Herbig stars with high-quality parallax measurements from GAIA DR2 \citep{GaiaADR22018}. In addition, 34 Herbig stars were classified as belonging to a low-quality parallax measurement sample. We take the stellar properties (mass, age) from \citet{Vioque2018}, who determine these from comparison to isochrones in the HR-diagram (see Table 1 of \citet{Vioque2018} for an overview of the values we adopted). The bottom panel of Fig. \ref{fig:Herbig_sample_hist} shows the distribution of a total of 252 Herbig stars over distance. The first peak at a distance of $\sim200$~pc consists of the nearby star-forming regions. The second peak at a distance of $\sim400$~pc is the Orion star-forming complex, after which ---at much larger distances--- the distribution goes up again due to a fast increase in volume. Due to incompleteness, the number of sources decreases again at large distances. From this figure, two natural cutoff  distances emerge to define our sample and a subsample of all Herbig disks. One cutoff  distance is at 450~pc. This distance includes the Orion star-forming complex and has a relatively good sampling of Herbig disks observed with ALMA compared to the full sample of disks given by \citet{Vioque2018}. After 450~pc, this sampling becomes quite sparse (see the bottom panel of Fig. \ref{fig:Herbig_sample_hist}). The other cutoff  distance is at 225~pc, which excludes Orion, but includes the nearby star-forming regions, and has a more complete sampling. Using a limit of 450~pc (225~pc) leaves a total of 96 (39) Herbig stars as listed by \citet{Vioque2018} of which 42 (25) have ALMA archival data.

While classified as Herbig stars by \citet{Vioque2018}, four objects with ALMA archival data are not included in our work: HD~143006 and BP~Psc are left out because they have a G5 and G9 spectral type, respectively \citep{Pecaut2016, Torres2006}, HD~135344 was not included as it is often mistaken for the B component of the binary \citep{Sitko2012}, and DK~Cha because it is in the Class I phase \citep{vanKempen2010, Spezzi2008}. In addition, the data of HD~36982 and HD~50138 do not satisfy the data selection criteria: for a proper comparison with other surveys, only ALMA Band 6 and 7 are used (see \S\ref{subsec:data_retrieval} for further details on the selection criteria). While being at the boundary of the definition of Herbig stars, we retained several objects in the interest of obtaining a  sample that is as complete as possible: HD~9672, HD~9672, Z~CMa, HD~58647, MWC~297, HD~53367, and R~CrA. For more details on the individual motivations, see Appendix \ref{app:details_boundary_objects}. Lastly, this work contains a few disks that would be classified as more evolved debris disks, especially among the group II objects (e.g., HD~9672 and HD~141569). However, all objects in our work do meet the definition of Herbig stars and are therefore kept in the sample.

In summary, of the 42 objects from \citet{Vioque2018} within 450~pc that have ALMA archival data, 36 data sets were used in this research (see the blue distribution in the top panel of Fig. \ref{fig:Herbig_sample_hist}). This set of 36 Herbig disks is the full sample used in this work. The 25 out of 36 disks that are within 225~pc are referred to as the nearby sample. All 36 objects are listed in Table \ref{tab:Herbig_params}. The used data sets and their corresponding properties can be found in Table \ref{tab:project_IDs}. Ultimately, the nearby sample (up to 225~pc) covers 64\%  (25/39 objects) of the total number of Herbig stars present within this distance, and 38\% (36/96) of the full sample (up to 450~pc).

The distances of the full sample range from 57~pc to 429~pc with a median of 160~pc. The range in stellar mass is fairly narrow: 1.47~$M_\odot$ to 3.87~$M_\odot$ with MWC~297 being an outlier with 14.5~$M_\odot$. The median stellar mass is $2.0$~$M_\odot$. The age is fairly advanced, ranging from 0.04~Myr (MWC~297) to 18.5~Myr (KK~Oph) with a median of 5.5~Myr. As HD~53367, R~CrA, and V892~Tau belong to the low-quality sample of \citet{Vioque2018}, their masses and ages were not determined. Although B stars are over-represented in the overall sample of \citet{Vioque2018}, our volume-limited sample is likely complete with respect to these spectral types and contains more A and F stars than B stars (especially early B stars, because none of these bright stars are likely missed at the close distances we are considering). The full sample is divided into 9~B-type, 20~A-type, and 7~F-type spectral type stars (see Table \ref{tab:Herbig_params} for references). As mentioned in \S\ref{sec:introduction}, Herbig Ae stars have horizontal PMS evolutionary tracks. Therefore, it is plausible that the sample used in this work consists of stars with relatively old ages. However, these selection effects are likely to be minimal both because the nearby sample is well covered and F-type stars are included. Lastly, there are 19~group~I and 16~group~II sources following the \citet{Meeus2001} classification (see Table \ref{tab:Herbig_params} for references). The Meeus group for HD~53367 is undetermined.


\section{Data and their reduction}
\label{sec:data_and_reduction}

\subsection{Data retrieval}
\label{subsec:data_retrieval}
Most of the data were obtained as product data from the ALMA archive\footnote{\url{https://almascience.eso.org/asax/}}. For the retrieval of the data, a few criteria were taken into account. First, the data have to be in Band 6 or Band 7 to both cover a similar wavelength range as previously done for population studies and to be able to compare the dust radii between Herbig disks and T~Tauri stars with data taken in different bands \citep[dust radii in Band 6 and 7 are comparable in size][]{Tazzari2021}. In addition, this minimizes the effects of dust evolution (grain growth and radial drift) that impacts different grain sizes differently; by using the same bands, we sample the same grain sizes. Second, the data with the highest resolution were often chosen ---except if the largest recoverable scale was too small--- to minimize the chance of resolving out large-scale structures. Lastly, if the first two criteria left multiple data sets to choose from, the data set with the highest continuum sensitivity was chosen.

All ALMA product data are imaged and calibrated using the standard ALMA pipeline, ensuring uniform data for all objects and thus well-suited product data for use in this work. However, some of the data were not directly available as a product. We therefore reduced the data of HD~37258, HD~176386, and TY~CrA ourselves. The reduction of all three data sets was done in the \texttt{Common Astronomy Software Applications} (CASA) application version 5.7.0 \citep{McMullin2007}. To increase the sensitivity of the continuum image and ultimately be able to adequately determine the dust mass of the disk, natural weighting was used for all three data sets. Because of a truncation error when setting the observation coordinates, the coordinates of HD~176386 and TY~CrA are shifted resulting in an offset of the sources of up to $15^{\prime\prime}$ east of the phase centre \citep{Cazzoletti2019}. The phase centre was therefore shifted to the correct source positions. For all three datasets, the signal-to-noise ratio (S/N) was deemed to be too low for self calibration to be applied. For the pipeline products, self calibration would likely have increased the S/N of the images, but would not significantly affect the total flux (which are all well above the S/N limit) and the inferred radii.

\begin{figure*}[b!]
    \centering
    \includegraphics[width=\textwidth]{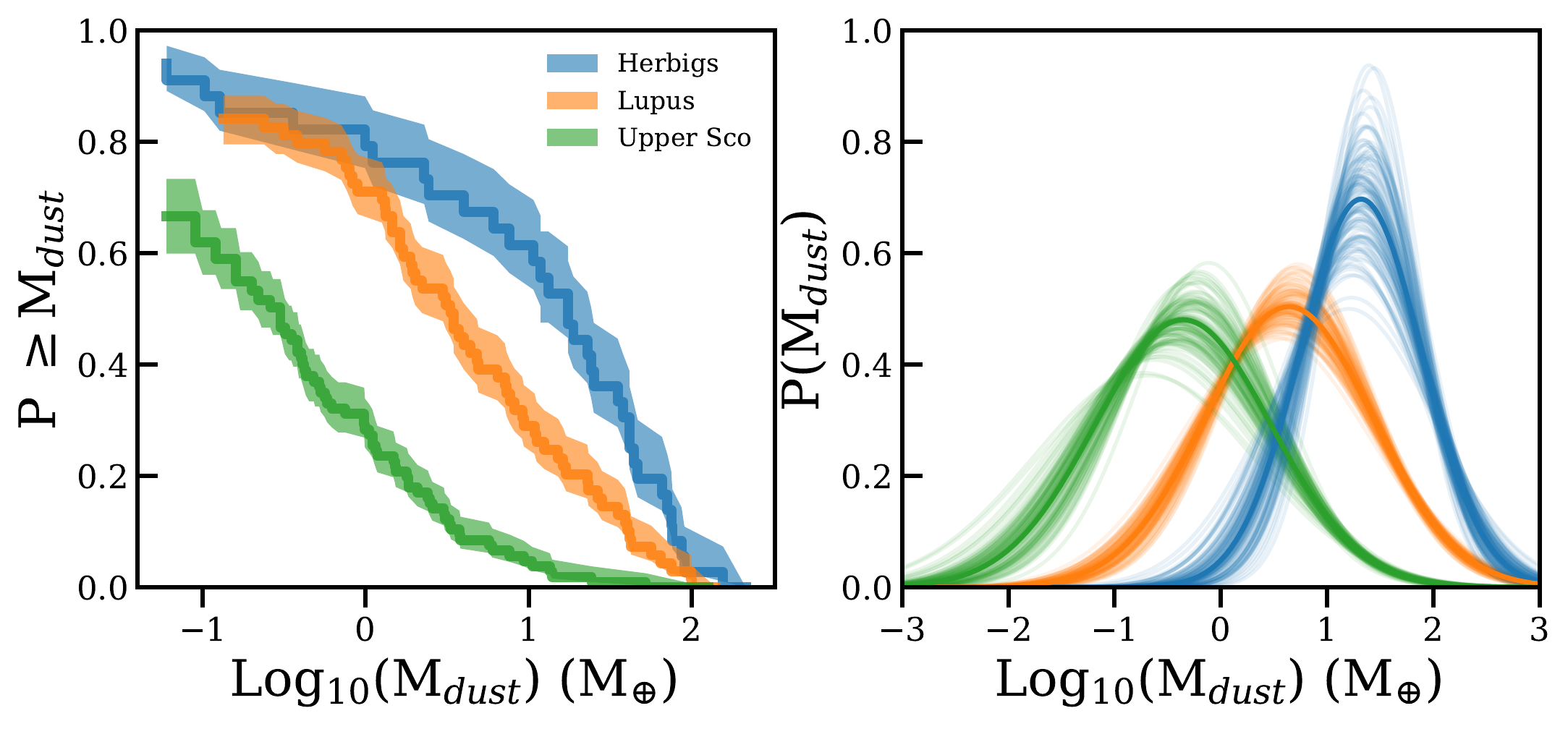}
    \caption{\textbf{Left:} Cumulative distribution functions of the dust masses contained inside the disks of our Herbig sample, Upper~Sco \citep{Barenfeld2016} and Lupus \citep{Ansdell2016}. \textbf{Right:} log-normal fit through the cumulative distributions. The solid line represents the best-fit distribution, while the light lines show a subsample of distributions from a bootstrapping method, showing the spread in possible fits.}
    \label{fig:CDFs}
\end{figure*}

Figure \ref{fig:gallery} shows the images of all detected disks. In general, the obtained set of data products consists of a large range of spatial resolutions. All images in which substructure is visible are considered to be resolved; for the remaining disks a Gaussian was fitted using the \texttt{CASA} \texttt{imfit} task. For all of these disks, the FWHM major axis found deviated by up to a few percent from the beam major axis and was therefore considered to be unresolved. Consequently, out of the 36 data sets, 17 are resolved. The resolution of the resolved data ranges from $0.02^{\prime\prime}$ to $1.7^{\prime\prime}$ with a median of $0.2^{\prime\prime}$. The resolution of the unresolved data ranges from $0.2^{\prime\prime}$ to $8^{\prime\prime}$ with a median of $0.7^{\prime\prime}$. Three data sets are taken in Band 7 (275--373~GHz), while the other data sets are taken in Band 6 (211--275~GHz). The sensitivity ranges from an rms of 0.02~mJy~beam$^{-1}$ to 12.4~mJy~beam$^{-1}$ with a median of 0.31~mJy~beam$^{-1}$. See Appendix \ref{app:archival_data_used} for specific information for each data set.
Lastly, the data of HD~53367 and HD~176386 are listed as non-detections, because continuum emission was only present at the $2.4\sigma$ and $1.6\sigma$ level respectively. For R~CrA no emission is detected and is listed as an upper limit as well.

\subsection{Determination of disk flux and size}
\label{subsec:flux_measurements}
The flux was measured using the same method as that adopted by \citet{Ansdell2016}, who use aperture photometry where the radius of the aperture is determined by a curve-of-growth method. An aperture with a specific position angle and inclination was centered on the disk. For the resolved disks, these were retrieved from other works; see Table \ref{tab:Herbig_params} for the values and references. For the unresolved disks, a circular aperture was used. The radius of this aperture was increased with a minimum step size set by the pixel size until the change in the total disk integrated flux by increasing the radius was smaller than five times the RMS noise. While the noise depends on the weighting scheme used, different robust parameters only minimally influenced the final found radius. This method gives the maximum disk-integrated flux emitted by the disk and in addition returns the radius of the disk. To facilitate comparison with the results of \citet{Ansdell2016}, we report $R_{90\%}$, the radius where 90\% of the total flux is reached. We also report $R_{68\%}$, where 68\% of the flux is reached, in order to allow comparison with other studies that report this value. Table \ref{tab:Herbig_params} lists the dust masses and the $68\%$ and $90\%$ radii.

For unresolved disks, we use the curve of growth method as well, as opposed to the Gaussian fitting method used by \citet{Ansdell2016}, because the difference between the two methods was found to be within 1\%. Considering that a standard 10\% flux calibration error has to be taken into account as well, these differences are deemed negligible.

\section{Results}
\label{sec:results}

\subsection{Dust masses}
The (sub)millimeter continuum flux can be directly related to the dust mass assuming optically thin emission \citep{Hildebrand1983} using

\begin{equation}
    M_\text{dust} = \frac{F_\nu d^2}{\kappa_\nu B_\nu\left( T_\text{dust} \right)},
    \label{eq:flux_to_dust}
\end{equation}

\noindent where $F_\nu$ is the continuum (sub)millimeter flux as emitted by the dust in the disk and $d$ the distance to the object. $B_\nu(T_\text{dust})$ is the value of the Planck function at a given dust temperature $T_\text{dust}$. In many population studies, the dust temperature is taken to be a single constant value across the disk with $T_\text{dust}=20$~K \citep[e.g.,][]{Ansdell2016, Cazzoletti2019} as found for the mean of the disks in the Taurus region \citep{Andrews2005}.

Given the high luminosity of the Herbig stars, we explicitly take the higher expected dust temperatures into account. We used the approach by \citet{Andrews2013} to estimate the dust temperatures,   where the mean dust temperature of the disk scales with the stellar luminosity as

\begin{equation}
    T_\text{dust} = 25\text{ K}\times\left( \frac{L_*}{L_\odot} \right)^{1/4}.
    \label{eq:Tdust_Lstar}
\end{equation}

\noindent As shown by \citet{Ballering2019}, taking $T_\text{dust}=20$~K as well as assuming a luminosity-scaled dust temperature via Eq. (\ref{eq:Tdust_Lstar}) can lead to underestimation of the dust temperature by $\sim10$~K. However, for Herbig disks, Eq. (\ref{eq:Tdust_Lstar}) leads to a less substantial underestimation of the temperature compared to the assumption of $T_\text{dust}=20$~K.

Following previous studies \citep[e.g.,][]{Ansdell2016, Cazzoletti2019}, the dust opacity at a given frequency $\kappa_\nu$ is given by a power law such that it equals 10~cm$^2$g$^{-1}$ at a frequency of 1000 GHz \citep{Beckwith1990} which is scaled with an index of $\beta=1$. While \citet{Tychoniec2020} found from combining VLA and ALMA data that the power-law index $\beta$ is $\sim0.5$, indicating growth of dust particles \citep[e.g.,][]{Natta2004, Ricci2010, Testi2014}, for the purpose of comparing with previous population studies, we use $\beta=1$ in this work.

Lastly, the errors on the dust masses as given in Table \ref{tab:Herbig_params} are calculated by taking two sources of error into account. First, the 10\% flux calibration and second, the error on the distance by averaging the upper and lower confidence intervals on the distance as given by \citet{Vioque2018}. While previous studies did not take the error on the distance into account because all the stars reside in the same SFR, it is important in this study because the Herbig stars are distributed over a large range of distances.

\subsection{The dust mass distribution}

\renewcommand{\arraystretch}{1.2}
\begin{table}[t!]
\caption{Log-normal distribution fit results for the dust mass cumulative distributions shown in Fig. \ref{fig:CDFs}. The $M_\text{dust}$ parameters are given in log$_{10}$($M/M_\oplus$).}
\centering
\begin{tabular}{l|cc|cc}
\hline\hline
          & \multicolumn{2}{c|}{$M_\star$ ($M_\odot$)} & \multicolumn{2}{c}{$M_\text{dust}$}   \\ \hline
          & $\mu$         & $\sigma$      & $\mu$                   & $\sigma$               \\ \hline
Herbigs   & 2.44          & 2.24          & 1.32$^{+0.05}_{-0.05}$  & 0.57$^{+0.07}_{-0.07}$ \\
Lupus     & 0.42          & 0.48          & 0.64$^{+0.04}_{-0.05}$  & 0.79$^{+0.05}_{-0.04}$ \\
Upper Sco & 0.43          & 0.37          & -0.36$^{+0.11}_{-0.14}$ & 0.83$^{+0.09}_{-0.07}$ \\ \hline
\end{tabular}\\
\label{tab:fit_params}
\end{table}

We derive the dust mass cumulative distribution function (CDF) using the \texttt{lifelines} Python package \citep{DavidsonPilon2021}. \texttt{Lifelines} uses survival analysis to construct a cumulative distribution of the dust masses, as shown in the left panel of Fig. \ref{fig:CDFs}. The \texttt{lifelines} package can take left-censored data into account (i.e., upper limits on the dust mass). The $1\sigma$ confidence intervals are shown as the vertical spread in the distribution, reflecting the number of samples in each bin. For a discussion on the completeness of the sample, see Appendix \ref{app:completeness_and_bias} where three extra Herbig disks are added from JCMT/SCUBA (sub)millimeter observations.

To understand the differences as a function of stellar mass, the dust mass CDF of the Herbig disks is compared in the left panel of Fig. \ref{fig:CDFs} with the dust mass CDFs of the young (1--3~Myr) Lupus \citep{Ansdell2016} as well as with the older (5--11~Myr) Upper~Sco \citep{Barenfeld2016} star-forming regions obtained by assuming a dust temperature of 20~K.

In general, using a luminosity-dependent $T_\text{dust}$ for the Lupus sample (with luminosities from \citealt{Alcala2014}) marginally changes the distribution, with the mean mass increasing by only 4\% and the median by 14\%. We therefore retain the $T_\text{dust}=20$~K derived masses for consistency. HR~5999 is the only object in \citet{Ansdell2016} for which a temperature of 20~K would not be appropriate. This Herbig disk is also in our sample and we infer a dust disk mass of 4.0~M$_\oplus$ instead of 23.3~M$_\oplus$ as reported by \citet{Ansdell2016}, because our determination uses $T_\text{dust}\approx67$~K which is more appropriate for an object of this luminosity (52~L$_\odot$).



The mean dust mass of the Herbig sample is $38\pm5$~M$_\oplus$. This is a factor of about three higher than the amount of dust in the disks in the Lupus star-forming region ($15\pm3$~M$_\oplus$) and a factor of about seven higher than the dust masses in the disks in Upper~Sco ($5\pm3$~M$_\oplus$). We note that while the assumption of optically thin emission may be valid for T~Tauri disks, this might not be the case for massive Herbig disks. However, if the Herbig disks are partially optically thick, the inferred mass from Eq. (\ref{eq:flux_to_dust}) would underestimate the true mass; correcting for this would even further increase the differences in dust mass between the Herbig sample and the Lupus and Upper~Sco star-forming regions. Our finding that Herbig disks have a higher mass is therefore robust. These higher dust masses are consistent with the well-known stellar mass ($M_\star$)--dust mass ($M_\text{dust}$) relation \citep[e.g.,][see \S\ref{subsec:Mstar_Mdust_relation} for further discussion on this]{Andrews2013, Barenfeld2016, Ansdell2016, Pascucci2016}. Following \citet{Williams2019}, the right panel of Fig. \ref{fig:CDFs} presents the result of fitting a log-normal distribution through the found cumulative distributions as shown in the left panel of Fig. \ref{fig:CDFs} (for the best-fit parameters, see Table \ref{tab:fit_params}). The resulting distributions provide a more straightforward depiction of the means and standard deviations of the obtained CDFs.  To account for the asymmetric errors on the CDFs obtained with the \texttt{lifelines} package, a bootstrapping method is used via sampling a split-normal distribution $10^5$ times, with the left and right standard deviation of the split-normal distribution given by the upper and lower errors on the CDF. The solid line in the right panel of Fig. \ref{fig:CDFs} is the best-fit log-normal distribution, while the fainter lines show the spread in possible fits. From Fig. \ref{fig:CDFs} it is clear that the Herbig disks indeed contain more dust mass than both Lupus and Upper~Sco. At the same time, the distribution of the older Upper~Sco region is lower (more than one standard deviation) than both the Lupus and Herbig disks, which is likely due to its advanced age (approximately one-third of the disks in Upper~Sco are more evolved class III objects \citealt{Michel2021}).

\begin{figure}[t!]
    \centering
    \includegraphics[width=0.5\textwidth]{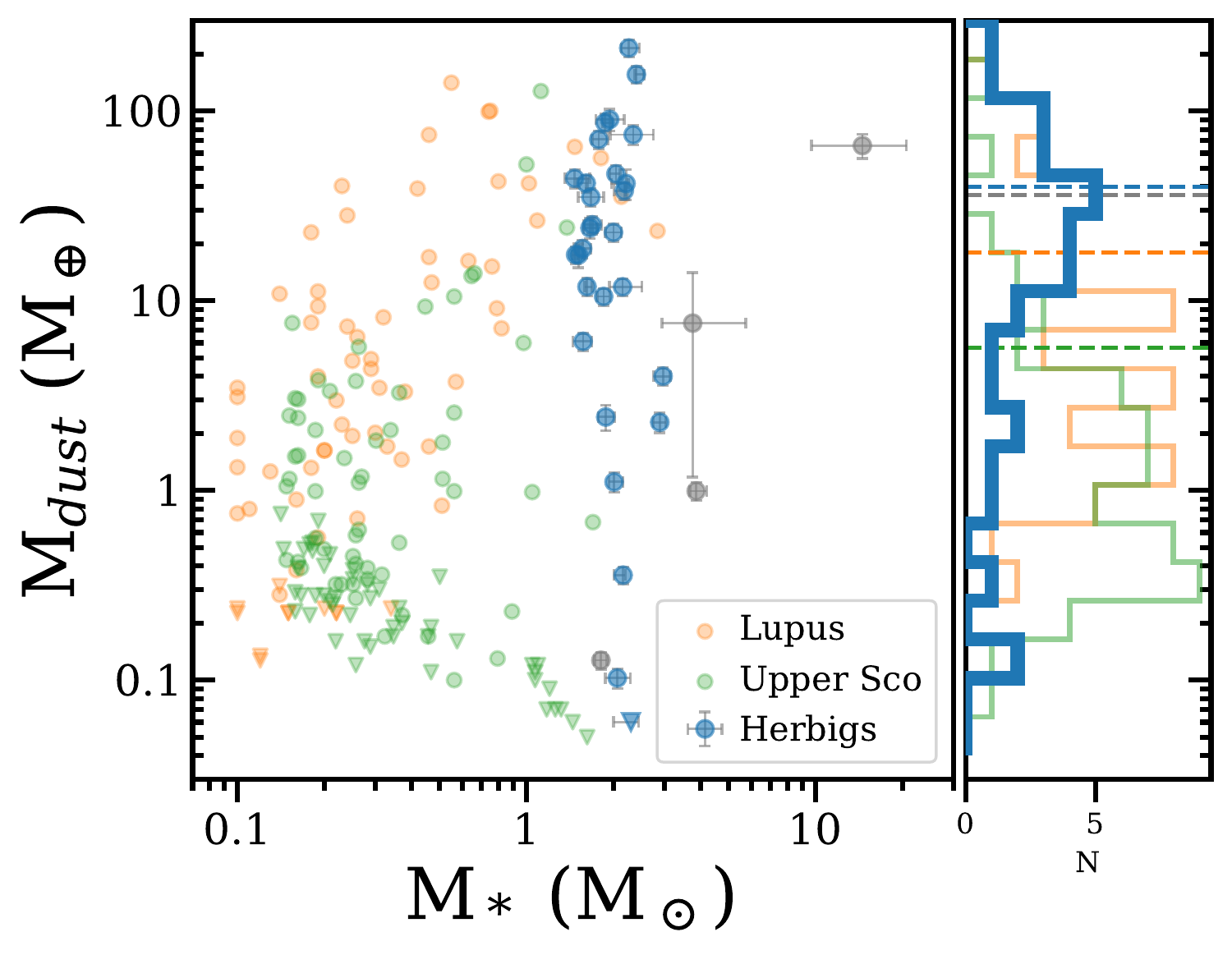}
    \caption{Stellar mass vs measured dust mass inside the disk. We show the dust masses of Upper~Sco in green \citep{Barenfeld2016}, Lupus in orange \citep{Ansdell2016}, and the ALMA Herbig sample in blue. On the right a histogram is plotted of the distribution of the dust masses. The grey points are four of the six borderline stars (see Sect. \ref{sec:sample_selection}). HD~53367 and R~CrA are missing because no stellar mass is available from \citet{Vioque2018} or \citet{Wichittanakom2020}.}
    \label{fig:Mstar_vs_Mdust}
\end{figure}

The now clearly identified relation between stellar mass ($M_\star$) and dust mass ($M_\text{dust}$) is consistent with this finding. Figure \ref{fig:Mstar_vs_Mdust} plots the stellar mass against the dust mass for the Herbig, Lupus, and Upper~Sco samples together with a histogram showing the distribution of the detected dust masses of each sample. While the mean dust mass of the Herbig disks is indeed higher than that of Lupus and Upper~Sco, this figure shows that the range in dust mass is relatively similar in the two low-mass star-forming regions. The highest dust masses of Lupus and Upper~Sco are $141\pm21$~M$_\oplus$ (Sz~82) and $127\pm31$~M$_\oplus$ (2MASS~J16113134--1838259) respectively, while the highest mass in the Herbig sample is $215\pm22$~M$_\oplus$ (HD~142527), which is a factor of 1.5 higher. The biggest difference is in the distribution of the masses. The Herbig sample contains many more massive disks than the low-stellar-mass sources. Of the Herbig sample, $\sim63$~\% of the disks have a mass of more than $10$~M$_\oplus$. Compared to only $\sim26$~\% of Lupus and $\sim6$~\% of Upper~Sco disks.

\begin{figure}[t!]
    \centering
    \includegraphics[width=0.5\textwidth]{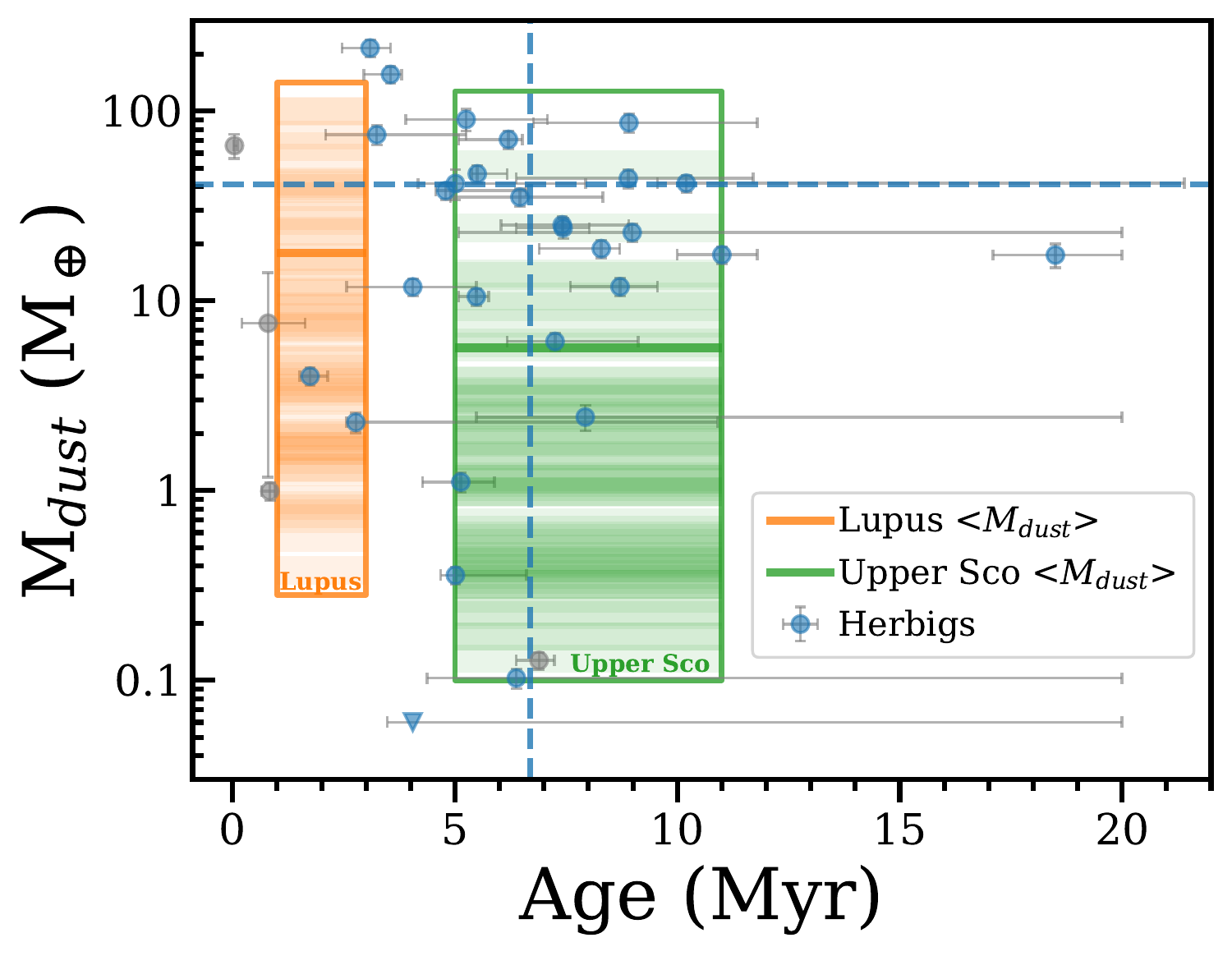}
    \caption{Disk dust mass vs the age of the star. The blue points are the ALMA Herbig sample and the grey points are the four out of six borderline stars (see Sect. \ref{sec:sample_selection}). The range in dust mass and age of the Lupus and Upper~Sco star-forming regions are shown as colored regions, where the gradient indicates the number of disks of a specific dust mass and the thick solid horizontal colored lines show the mean dust masses. The blue horizontal and vertical dashed lines show the respective mean values of the Herbig sample.}
    \label{fig:Age_vs_Mdust}
\end{figure}

Figure \ref{fig:Age_vs_Mdust} presents the age of the disks plotted against the dust mass. The ranges in detected dust masses and ages of the Lupus and Upper~Sco SFRs are shown as well. The mean dust mass of the Herbig disks, shown by the horizontal dashed line, is higher than the mean dust masses of both Lupus and Upper-Sco, as shown by the solid horizontal lines. Additionally, this plot shows that the mean age of the Herbig sample mostly overlaps with the Upper~Sco region. This suggests that Herbig disks either retain their disk mass for longer, and/or that they are initially formed with higher dust masses following the $M_\star$--$M_\text{dust}$ relation and thus still have a large disk mass after a few million years. This is discussed further in \S\ref{subsec:Mstar_Mdust_relation}. Additionally, when removing the six borderline objects from the sample, the distribution as shown in the left panel of Fig. \ref{fig:CDFs} does not change significantly. It is therefore unlikely that these targets are much different from the other Herbig disks.

\subsection{Herbig disk dust radii}
We also determined the Herbig disk dust radii at 68\% ($R_{68\%}$) and 90\% ($R_{90\%}$) of the total flux (see Table \ref{tab:Herbig_params}). Based on these measurements, the majority of the Herbig disks seem to be large compared to T~Tauri disks. For the resolved disks, R$_{68\%}$ ranges from 35±2~au (HD~100546) to 315±39~au (HD~9672), with a median radius of 71±7~au (HD~169142). The minimum and maximum radius of R$_{90\%}$ is slightly larger at 41±4~au and 473±112~au for the same disks, with a median radius of 109±31~au (HD~135344B). We should stress that this is the \emph{dust} radius and the \emph{gas} radius is most likely a factor of a few larger \citep[e.g.,][]{Ansdell2018, Trapman2019, Sanchis2021}. Also, it is possible that there is some tenuous dust emission at larger radii, which was for example found for HD~100546 \citep{Walsh2014, Fedele2021}. But this would not strongly influence $R_{90\%}$.

\begin{figure}[t!]
    \centering
    \includegraphics[width=0.5\textwidth]{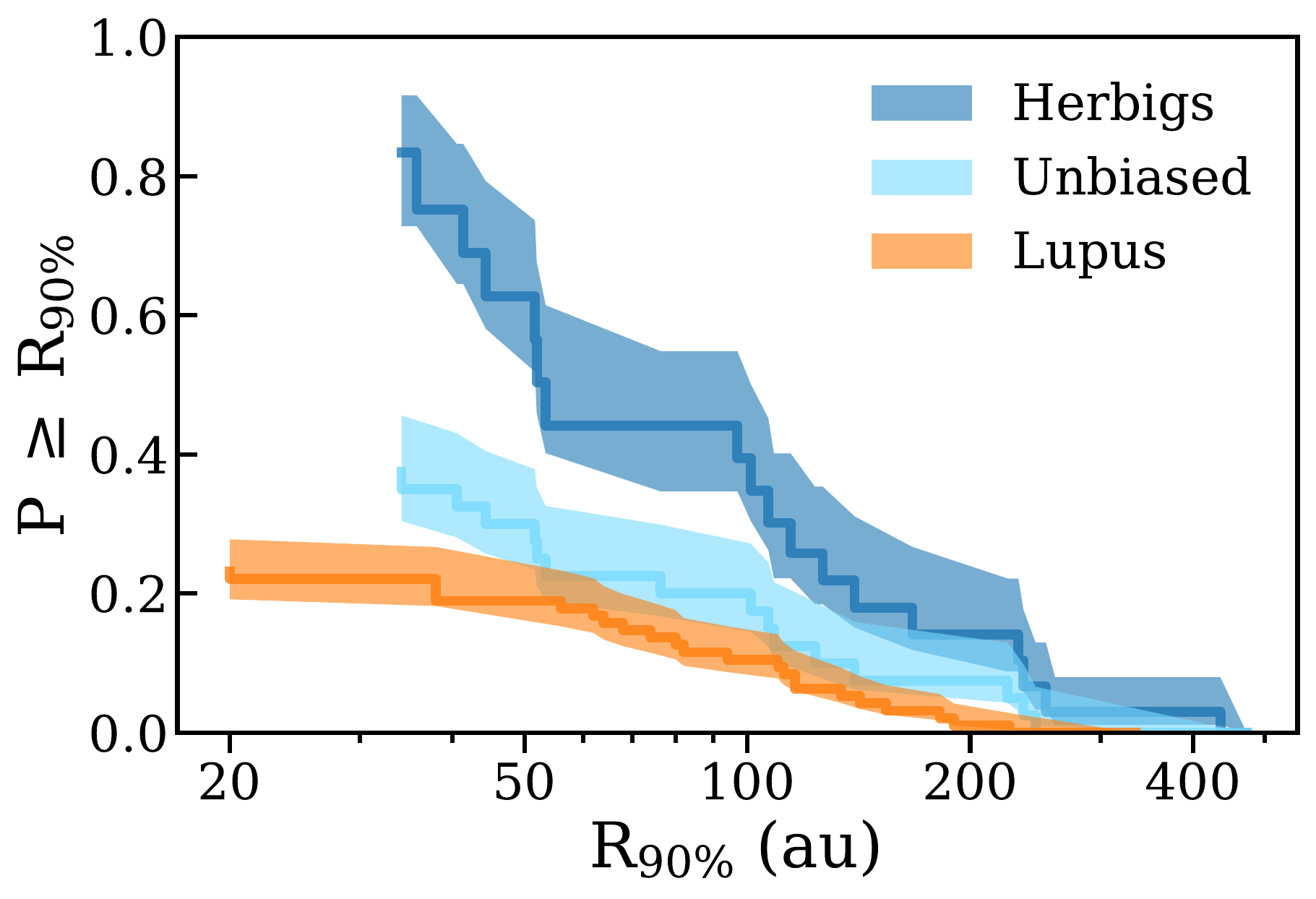}
    \caption{Cumulative distribution functions of the 90\% dust radii of the Herbig and Lupus samples \citep{Ansdell2018}. The unbiased distribution is made by assuming that all unresolved and unobserved disks are smaller than or equal in size to the smallest disk present in the sample.}
    \label{fig:CDF_radii}
\end{figure}

The larger confidence intervals of R$_{90\%}$ are due to the curve-of-growth levelling off, making the range in which the flux is within its 10\% error rather large. The errors on R$_{68\%}$ are on the other hand set by the discrete number of pixels in the image, where the pixel size is dictated by the resolution (synthesized beam). For the unresolved disks, the radius is constrained by the computed upper limits. The R$_{68\%}$ upper limits range from 27~au (TY~CrA) up to 328~au (V599~Ori) and have a median of 82~au. The R$_{90\%}$ upper limits range from 34~au (TY~CrA) up to 326~au (HD~37258) and have a median of 126~au (HD~245185).

For Lupus, 76\% of all detected disks were {unresolved} \citep{Ansdell2018}. Using half of the FWHM beam size ($0.25^{\prime\prime}/2$) and a mean distance of 160~pc \citep{Manara2018}, the upper limit on R$_{90\%}$ for these unresolved disks is <20~au. Consequently, the mean R$_{90\%}$ dust radius is at most 42~au and the median at most 20~au. This is a factor of about three smaller compared to the median radius of the Herbig disks. For Upper~Sco, the same is found: \citet{Barenfeld2017} measured the dust radii of Upper~Sco by fitting power-law models to the dust surface density. They do not report $R_{90\%}$ radii, and so we compare their $R_{68\%}$ radii that have a median of 14~au and a range of 5~au to 44~au with our median Herbig $R_{68\%}$ dust radii. The latter are found to be a factor of about five larger.

These low mean dust disk radii do not mean that no large T~Tauri disks exist. Some T~Tauri dust disks in Lupus do extend out to R$_{90\%}=334$~au. This is clear in Fig. \ref{fig:CDF_radii}, which shows the cumulative distributions of the 90\% flux cutoff  radii of the Lupus and Herbig populations. For both distributions, the upper limits on the unresolved disks are taken into account. The range in dust radii is similar in both T~Tauri (the Lupus sample) and Herbig disks, but the large number of small T~Tauri disks skews the distribution to smaller radii compared to the Herbig sample. However, the distribution for the Herbig disks is an optimistic case because the unresolved disks are likely smaller than what is suggested by their upper limits, and therefore the distribution lies higher than it would if all disks were resolved. One can `unbias' the distribution by assuming all unresolved and unobserved disks to have the same upper limit as the smallest disk. The 225~pc limited sample is used, as this one is the most complete. This unbiased distribution is shown in light blue. Using a two-sample Kolmogorov-Smirnov test from the \texttt{Scipy} package \citep{2020SciPy-NMeth}, we can reject the null hypothesis of the Lupus and Herbig dust radii to be drawn from the same underlying distribution based on a $p$-value of $9.5\times10^{-5}$. The unbiased distribution is found to be similar to the Lupus distribution because the same null hypothesis cannot formally be rejected based on a $p$-value of $5.6\times10^{-2}$. Changing to R$_{68\%}$ (with the radii of Lupus from \citealt{Sanchis2021}), we obtain similar results. Hence, we conclude that the measured Herbig dust radii are significantly larger than those of Lupus.  The Herbig disk size distribution only becomes comparable to that in Lupus if all unresolved and all unobserved disks turn out to be as small as (or smaller than) 27~au (although Lupus also has undetected disks, which might be small, which are not taken into account here); in all other cases, Herbig disks are significantly larger than the disks in Lupus.

\section{Discussion}
\label{sec:discussion}

\subsection{The $M_\star$--$M_\text{dust}$ relation}
\label{subsec:Mstar_Mdust_relation}
The $M_\star$--$M_\text{dust}$ relation is one of the important conclusions from population studies in the past decade. By now, this relation is well established across a large variety of star-forming regions mostly containing lower mass stars \citep[e.g.,][]{Andrews2013, Ansdell2016, Ansdell2017, Pascucci2016, Barenfeld2016}. Our observations are in line with the increase in dust mass for increasing stellar mass. The exact relation does however depend on the considered stellar mass range. For the full sample used in this work, the range in stellar masses is rather narrow: 1.5~M$_\odot$ to 3.9~M$_\odot$, excluding MWC~297 with a mass of 14.5~$M_\odot$. As Fig. \ref{fig:Mstar_vs_Mdust} shows, among this small range of stellar masses, there is a large spread in dust mass. This might be due to the diversity in age and birth environment present in the sample, which changes the amount of dust present in the disk and complicates the comparison in terms of dust mass between the different distributions; this is further addressed in \S\ref{subsec:scaling_Mdust}. Both the narrow range in stellar mass and the diversity in age and birth environment hide any trend which could be present within the sample. However, a general trend is visible when considered in a wider stellar mass range (i.e., combined with Lupus and Upper~Sco); see the histograms in Fig. \ref{fig:Mstar_vs_Mdust}. The Herbig disks are more massive compared to the disks in Lupus and Upper~Sco.

\begin{figure}[t!]
    \centering
    \includegraphics[width=0.5\textwidth]{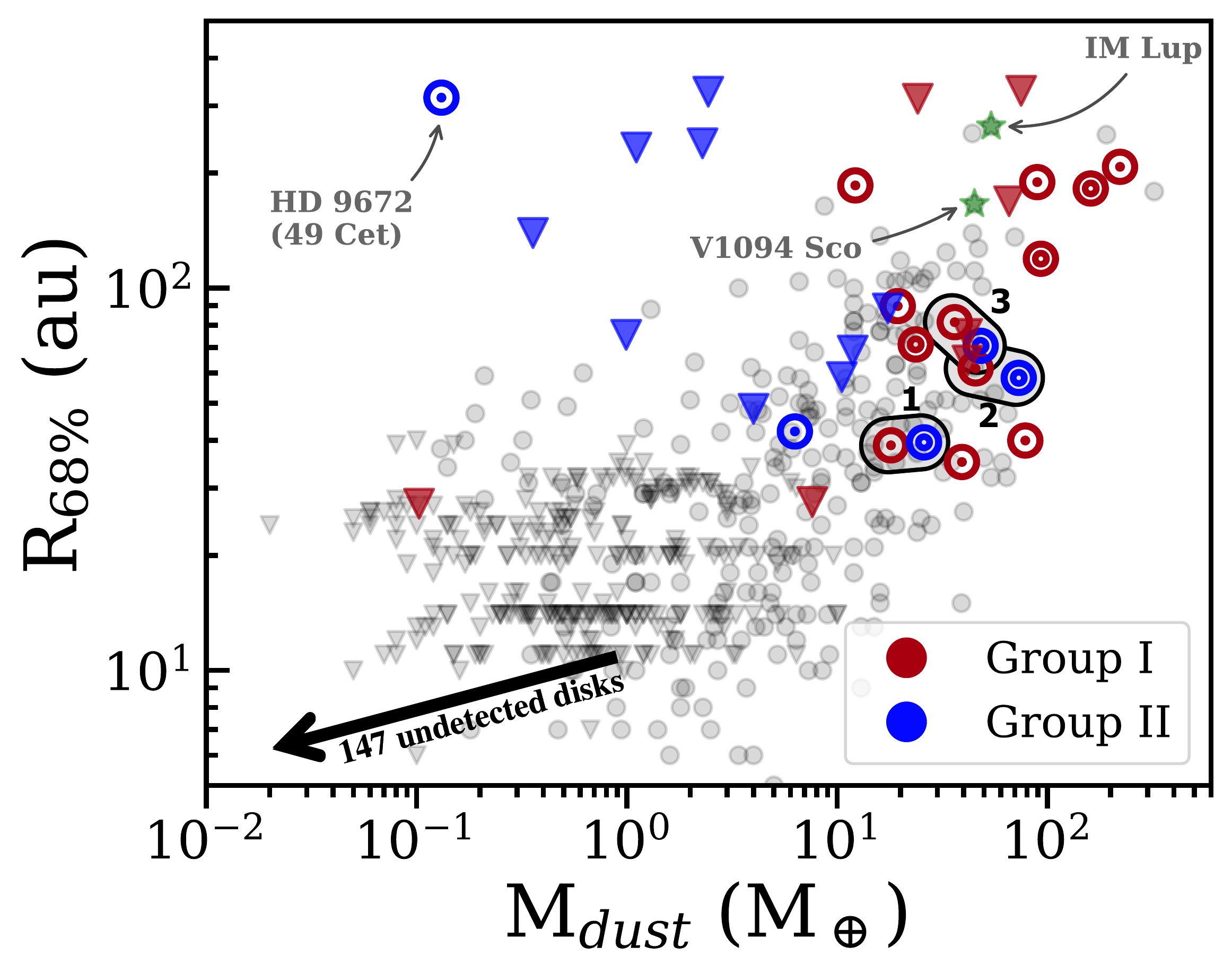}
    \caption{ R$_{68\%}$ radius plotted against the dust mass for the full sample in this work. Unlike Fig. \ref{fig:CDF_radii}, we use R$_{68\%}$ values here for consistency with the work of \citet{vanderMarel2021}. The Herbig disks are divided into the \citet{Meeus2001} group I and II disks. disks with one ($\odot$) and two or more ($\circledcirc$) (visible) rings are indicated as well. The upper limits on the radius are plotted as triangles. The grey scatter points and radius upper limits are the T~Tauri disks of \citet{vanderMarel2021}. The large arrow shows the range of upper limits on the undetected disks. Three pairs of Herbig group I and II disks are circled and numbered 1-3; these are further discussed in Sect. \ref{subsec:g1_vs_g2} and Fig. \ref{fig:g1_vs_g2}. For extra context, the positions of the large T~Tauri disks IM~Lup and V1094~Sco are shown as green stars \citep{Ansdell2016, Ansdell2018, vanTerwisga2018, Cleeves2016}.}
    \label{fig:meeus_groups_overview}
\end{figure}

The $M_\star$--$M_\text{dust}$ relation has also been found to steepen over time, mainly based on low-mass stars \citep[e.g.,][]{Pascucci2016, Ansdell2017}. The fact that the more massive disks seem to retain their disk mass for longer has been hypothesized to originate from inner holes and/or structures, which tend to be larger in more massive disks. \citet{GuzmanDiaz2021} found that $\sim28\%$ of Herbig Ae/Be disks can be considered to be transitional based on their SED (though uncertain because it can be affected by for example inclination and extinction; half of transition disks in Lupus show no signature of a cavity in the SED \citealt{vanderMarel2018}), which is more than the fraction of transitional disks in T~Tauri stars as imaged with ALMA \citep{vanderMarel2018}. Also, $\sim34\%$ of Herbig Be stars tend to have larger inner disk holes, significantly higher compared to $\sim15\%$ for Herbig Ae stars \citep[][who estimated the inner disk hole sizes from the SED, i.e., sizes from 0.1~au out to 10~au, which samples smaller radii than the large cavities with ALMA at tens to hundreds of au]{GuzmanDiaz2021}. This is in accordance with an observed increase in structure with stellar mass \citep{vanderMarel2021}. In our work there are two B-type stars with disks that have a dust mass higher than 10~M$_\oplus$. HD~34282 has a visible ring structure present in its disk, while R~CrA has been observed with the largest beam size of the data used in this work so that no structures are visible. For the later spectral types, all resolved disks indeed show disk structures.

Figure \ref{fig:meeus_groups_overview} shows the dust masses and dust radii for all detected disks in this work and indicates whether the disk is unresolved or a ring  ($\odot$) or multiple rings  ($\circledcirc$) are visible. This plot shows that in general, the disks with visible structure are more massive with a mean of $57\pm9$~M$_\oplus$ compared to the unresolved disks with a mean of $19\pm3$~M$_\oplus$, a factor of two lower. The unresolved disks mainly consist of the Herbig disks coming from population studies by for example \citet[][]{Cazzoletti2019}. These disks are expected to be the least biased towards large and bright disks. It is therefore not unlikely that the 17 resolved disks ($\sim47\%$ of the 36 disks) are more massive than the average Herbig disk. Still, the 225~pc limited sample contains 15 of the 17 resolved disks (HD 34282 and HD 290764 are at larger distances), all showing structure. This means that at least $\sim60\%$ (15/25) of the disks likely contain structure. This is an important result, as \citet{vanderMarel2021} show that there is a correlation between stellar mass and the presence of structure. These latter authors find that $\sim63\%$ of disks around stars with masses $>1.5M_\odot$ are structured (i.e., transition and ringed disks), which agrees well with our findings. Additionally, the more massive disks (i.e., $\gtrsim 10$~M$_\oplus$) seem to be more extended and show more structure than lower mass disks \citep{vanderMarel2021}. This also agrees with our findings here (which include a factor of about two more Herbig disks): the number of high-mass disks (64\% (23/36) of the disks have $\gtrsim$10~M$_\oplus$) does fall in line with the number of disks in which we see structure (15/25, $\sim60\%$).



 \begin{figure*}[t!]
    \centering
    \includegraphics[width=\textwidth]{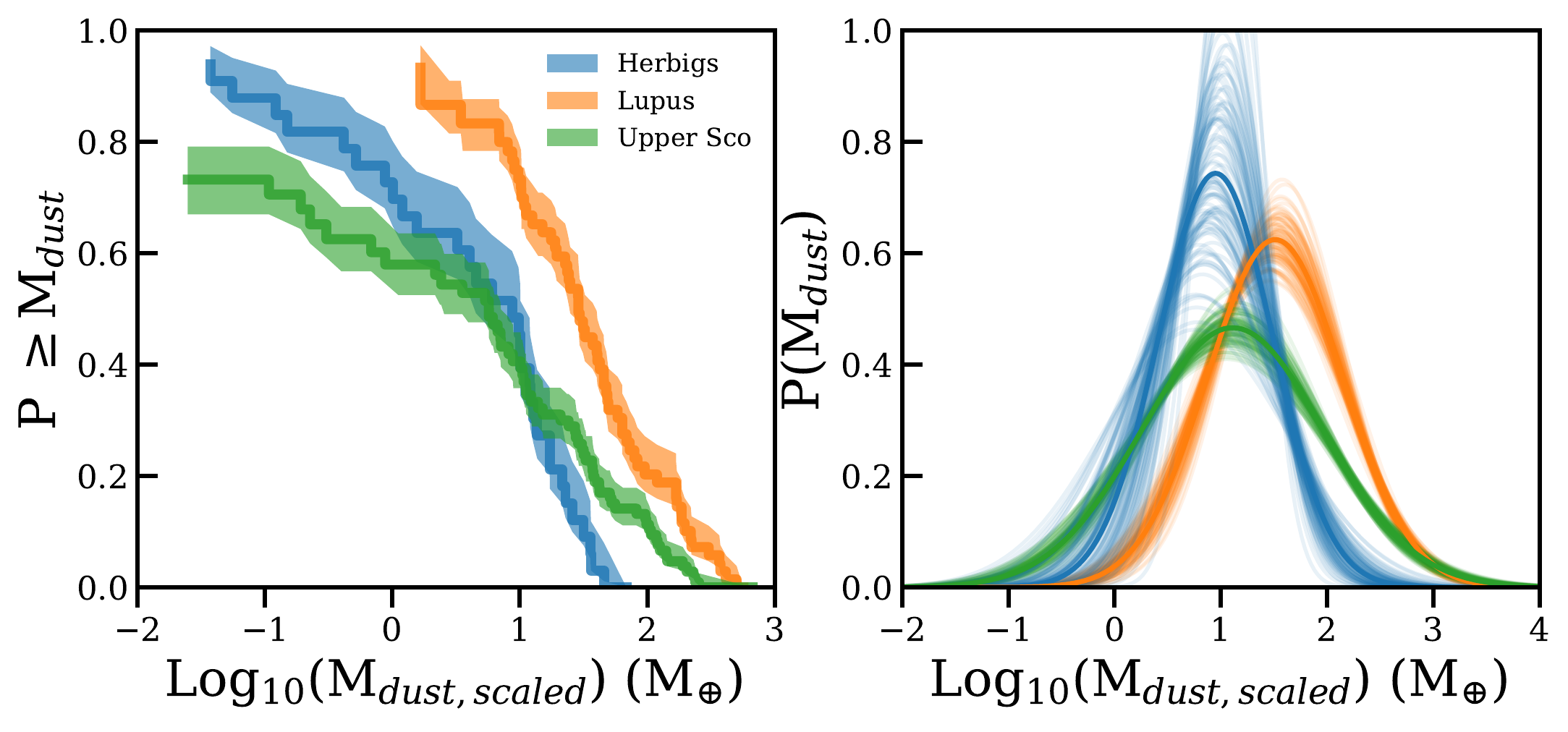}
    \caption{\textbf{Left:} Cumulative distributions of the dust masses scaled with the $M_\star$--$M_\text{dust}$ relation to a 1$M_\odot$ star. \textbf{Right:} log-normal fit through the cumulative distributions. The solid line represents the best-fit distribution, while the light lines show a subsample of distributions from a bootstrapping method, showing the spread in possible fits.}
    \label{fig:CDF_to_PDF_scaled}
\end{figure*}

\subsection{Scaling dust masses with $M_\star$}
\label{subsec:scaling_Mdust}
How does the Herbig disk mass distribution look if we remove the underlying $M_\star$--$M_\text{dust}$ dependency calibrated primarily at lower $M_\star$? To obtain an $M_\star$ independent mass estimate, the dust masses are scaled with the $M_\star$--$M_\text{dust}$ relation to the expected dust masses for a 1$M_\odot$ star using

\begin{equation}
    M_\text{dust, scaled} = \left( \frac{M_\star}{1 M_\odot} \right)^{-\alpha} \times M_\text{dust},
\end{equation}

\noindent where $\alpha$ is the slope of the power-law and $M_\star$ is in solar masses. The resulting CDFs can be found in the left panel of Fig. \ref{fig:CDF_to_PDF_scaled}. For both Lupus and Upper~Sco, the relations found by \citet{Ansdell2017} are used (respectively $\alpha=1.8$ and $\alpha=2.4$). For the Herbig sample, the relation found by \citet{Andrews2013} is used ($\alpha=1.4$) because this study includes more Herbig stars compared to other studies (see Table \ref{tab:population_studies}). Scaling the dust masses removes the effect that the stellar mass has on the dust mass and mostly leaves the effects of age considering that dust temperature was already accounted for with Eq. (\ref{eq:Tdust_Lstar}) for the Herbig disks and a temperature of 20~K for the low-mass star-forming regions. Again, following the same approach as \citet{Williams2019}, a log-normal distribution is fitted through the found CDFs; the fitted parameters are shown in Table \ref{tab:fit_params_scaled}. The best-fit log-normal distribution is shown in the right panel of Fig. \ref{fig:CDF_to_PDF_scaled}.

\begin{table}[b!]
\caption{Log-normal distribution fit results for the scaled dust mass cumulative distributions shown in Fig. \ref{fig:CDF_to_PDF_scaled}. The $M_\text{dust}$ parameters are given in log$_{10}$($M/M_\oplus$).}
\centering
\begin{tabular}{l|cc|cc}
\hline\hline
          & \multicolumn{2}{c|}{$M_\star$ ($M_\odot$)} & \multicolumn{2}{c}{$M_\text{dust, scaled}$}   \\ \hline
          & $\mu$         & $\sigma$      & $\mu$                  & $\sigma$               \\ \hline
Herbigs   & 2.44          & 2.24          & 0.95$^{+0.07}_{-0.09}$ & 0.54$^{+0.12}_{-0.11}$ \\
Lupus     & 0.42          & 0.48          & 1.51$^{+0.04}_{-0.04}$ & 0.64$^{+0.04}_{-0.04}$ \\
Upper Sco & 0.43          & 0.37          & 1.11$^{+0.04}_{-0.04}$ & 0.86$^{+0.05}_{-0.05}$ \\ \hline
\end{tabular}\\
\label{tab:fit_params_scaled}
\end{table}

For these scaled masses, the cumulative distribution of the Herbig disks moved towards lower dust masses because the stellar mass is above one solar mass. On the contrary, both Upper~Sco and Lupus moved towards higher dust masses. The width of both the Upper~Sco and Herbig disk distributions stayed similar (though the width of the Herbig disk distribution is not well constrained) because of the relatively narrow stellar mass distributions. Lupus has in contrast a broader stellar mass distribution, resulting in a narrower distribution after scaling the dust masses (see Table \ref{tab:fit_params}).

Applying the $M_\star$--$M_\text{dust}$ relation unsurprisingly moves the mass distribution of the Herbig disks to the same values as found in Upper~Sco and only slightly below the distribution found in Lupus. This scaling shows that the mass difference between the Herbig disks and the T~Tauri disks can indeed be tied to a difference in stellar mass. Also, the larger disk sizes we find for the Herbig disks may hold further clues as to why this relation exists.



\subsection{The $R_\text{dust}$--$M_\text{dust}$ relation}
\label{subsec:Rdust_Mdust}
Following \citet{vanderMarel2021}, we speculate that the larger disk masses and sizes found for Herbig disks as compared to their T~Tauri counterparts reflect the combined effect of inheritance and evolution. Our ansatz is that these findings originate from an initial disk mass that increases with stellar mass, and that subsequent disk evolution enlarges the differences as explained below, especially as these are obtained via (sub)millimeter continuum flux observations. It is a reasonable assumption that more massive cloud cores form more massive stars that are surrounded by more massive disks, if a constant fraction of the accreting material ends up in the disk. Although object-to-object variations will undoubtedly introduce a scatter around this relation, an overall scaling between the mass of the cloud core, the final mass of the star, and the mass of the disk is plausible.


A higher mass disk contains more material available for planet formation, and such planets can grow out to larger masses \citep[e.g.,][]{Mordasini2012}. This matches the observed trend that higher mass stars (1~M$_\odot$ and higher) are more often found to host planets more massive than $100$~M$_\oplus$ compared to stars with lower stellar masses \citep[e.g.,][]{Johnson2007, Johnson2010, Wittenmyer2020, Fulton2021}. Such planets can also be formed on larger orbits, because the required surface density is present at larger radii in a higher mass disk. This matches the presence of wide-orbit gas-giants around stars like HR~8799 \citep{Marois2008, Marois2010} and $\beta$~Pictoris \citep{Lagrange2010}, and (proto)planets in the disks like the ones around HD~100546 \citep{Quanz2013, Quanz2015} or HD~163296 \citep{Isella2016, Zhang2018, Liu2018, Teague2018, Pinte2018}. Once planets are formed of sufficient mass, gaps can open up, as is commonly observed in many disks, and inward transport of material, especially dust, is halted \citep[e.g.,][]{Zhu2012}. This has two observable effects. First, it traps a detectable mass reservoir outside the gaps that leads to larger dust continuum disk sizes. Second, by preventing the dust from flowing inward, a larger fraction of the dust remains in outer disk regions, thus allowing a larger continuum flux (because the more spread out material remains optically thin, and so all dust mass contributes, or because optically thick material subtends a larger region on the sky) and larger inferred disk dust mass \citep{Ballering2019}. A relation has indeed been found between the continuum size and the luminosity in disks \citep{Tripathi2017} and more recently, \citet{vanderMarel2021} found a general increase in dust mass with dust extent. Figure \ref{fig:meeus_groups_overview} presents the R$_{68\%}$ and M$_\text{dust}$ of the full sample of our work together with the 700 disks analyzed in \citet{vanderMarel2021}. The sample of \citet{vanderMarel2021} mainly contains T~Tauri disks, but does have $\sim50\%$ of the Herbig disks which are in our work. Figure \ref{fig:meeus_groups_overview} clearly shows that in general the dust disk size and mass of Herbig disks are larger than those of the typical T~Tauri disk, especially because $\sim21\%$ of the T Tauri disks are not detected.

\begin{figure*}[b!]
    \centering
    \includegraphics[width=\textwidth]{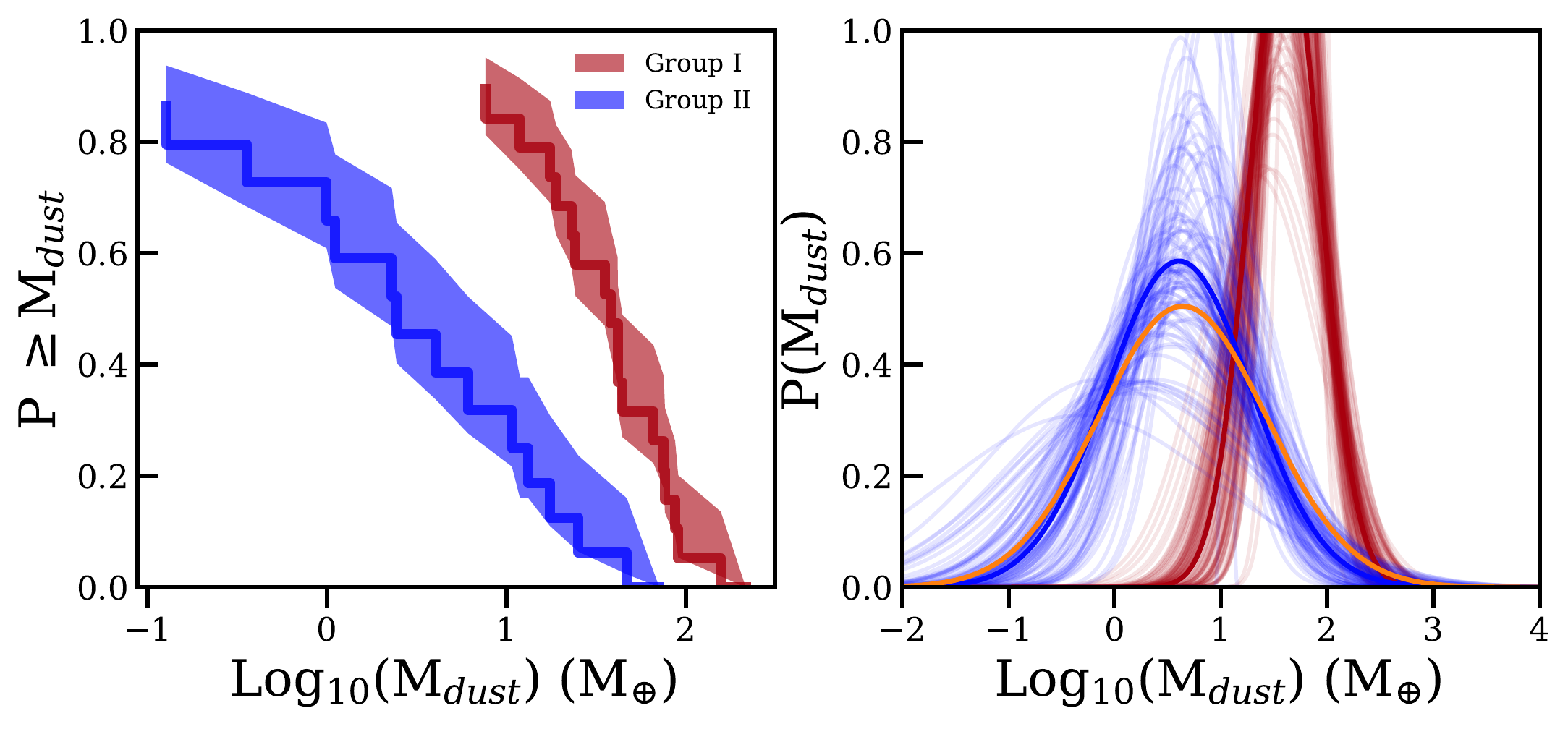}
    \caption{\textbf{Left:} Cumulative distributions of the dust masses of the group I and group II sources. \textbf{Right:}  log-normal fit through the cumulative distributions. The solid line represents the best-fit distribution, while the light lines show a subsample of distributions from a bootstrapping method, showing the spread in possible fits. The orange log-normal distribution shows the best-fit distribution of Lupus shown in Fig. \ref{fig:CDFs}.}
    \label{fig:CDF_to_PDF_meeus}
\end{figure*}

However, this does not mean that T~Tauri stars cannot have disks that are large and massive: there are T~Tauri disks that are as large and massive as a Herbig disk, but they make up a much smaller fraction of the total population. Figure \ref{fig:meeus_groups_overview} also shows the two largest disks in Lupus (V1094~Sco, K6-type, \citealt{vanTerwisga2018}; IM~Lup, M0-type, \citealt{Cleeves2016}), which are of similar size, but have a six times lower mass than the Herbig disks and are only 2\% of the Lupus population. The mass difference may simply reflect the dust mass distribution of Herbig disks peaking at masses a factor of about 6 higher than the dust mass distribution for T~Tauri stars.

HD~9672 (49~Ceti) is unusually large for its mass and a clear outlier in Fig. \ref{fig:meeus_groups_overview}. Its close distance (57~pc) has helped in detecting this low-mass source and identifying it as an outlier. This raises the question of whether other undetected objects may be harboring extended, low-surface-brightness disks.



\begin{figure*}[t!]
    \centering
    \includegraphics[width=\textwidth]{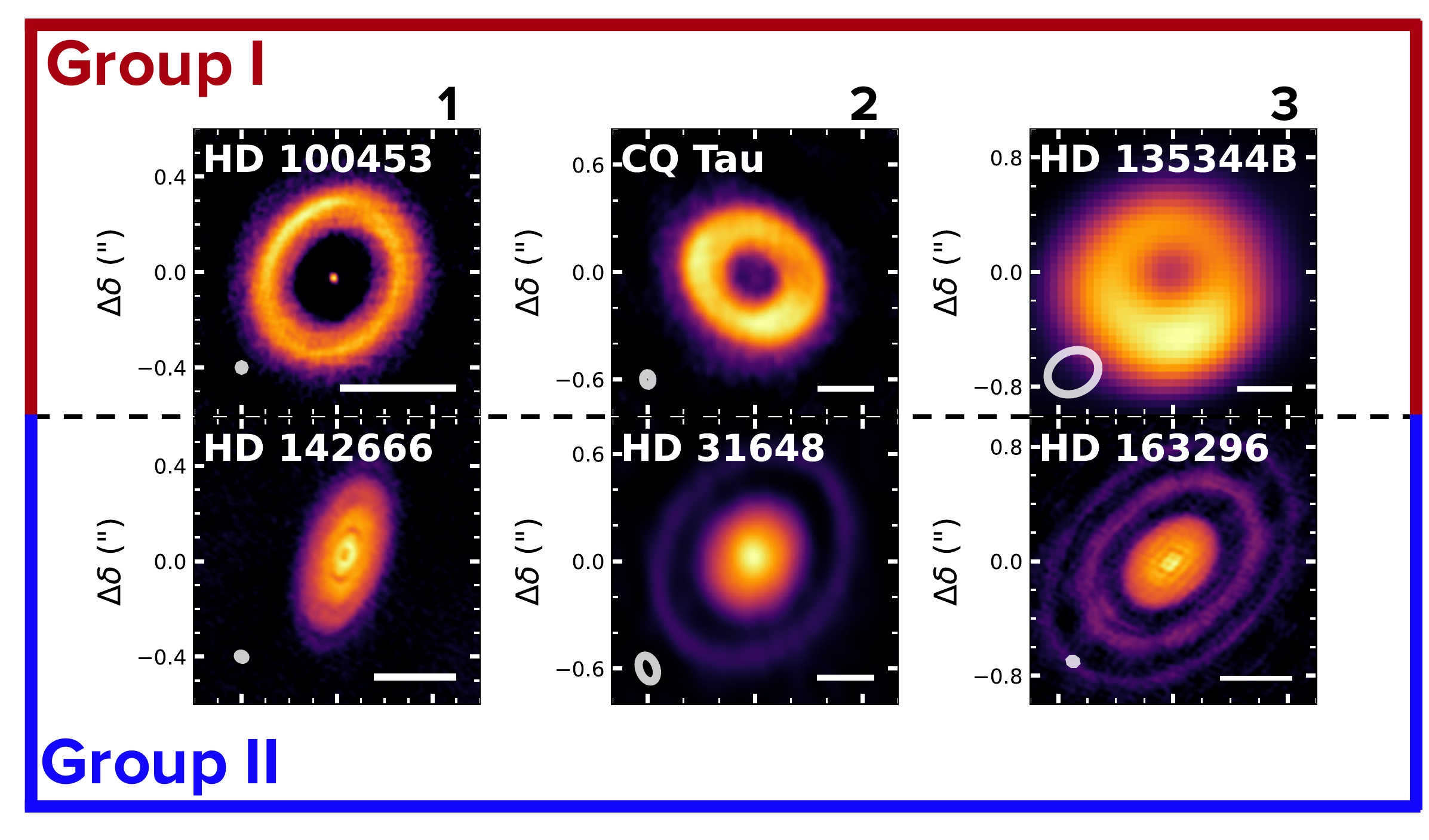}
    \caption{Three resolved disks for both group I and group II with similar extent and mass for each pair of disks. Each image is normalized with an asinh stretch to increase the visibility of the fainter details of the disk. A 50~au scale bar is shown to the bottom right of each image.}
    \label{fig:g1_vs_g2}
\end{figure*}

\begin{figure}
    \centering
    \includegraphics[width=0.5\textwidth]{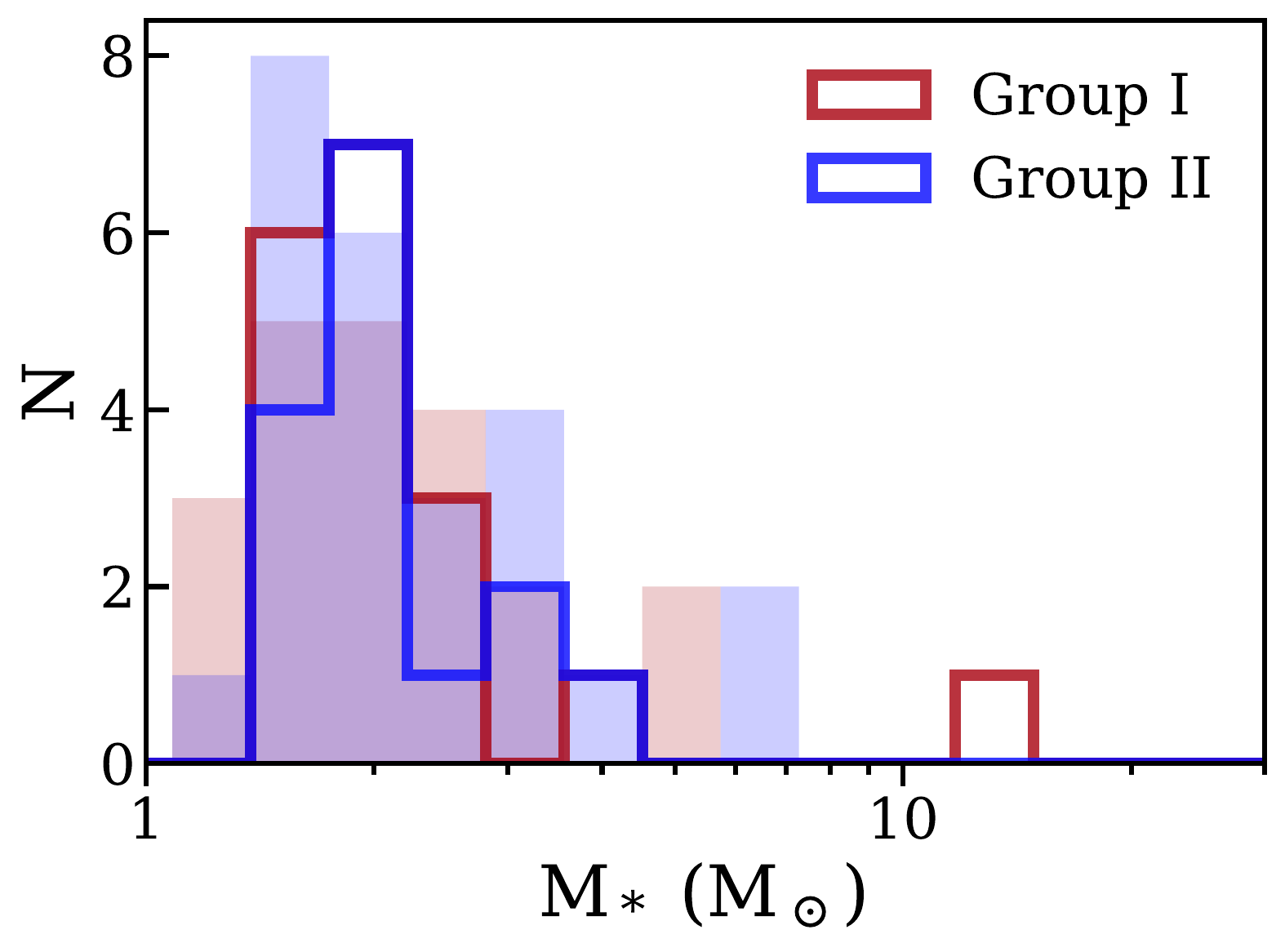}
    \caption{Stellar mass distribution of the group I and II objects used in this work. The shaded histogram shows the distribution of stellar masses for both the group I and group II objects not observed with ALMA within 450~pc.}
    \label{fig:groups_vs_mass}
\end{figure}

In contrast, if most or all of the dust has migrated radially inward to spatially unresolved disk regions, smaller disk sizes are inferred and, because inner disk regions will more easily become optically thick at (sub)millimeter wavelengths (for the same mass), lower continuum fluxes lead to lower inferred disk dust masses following Eq. (\ref{eq:flux_to_dust}). This is in line with the findings of \citet{Ballering2019}, who show that the inferred disk mass from Eq. (\ref{eq:flux_to_dust}) of T~Tauri disks in Taurus could underestimate the mass by a median factor of  about two. Neglecting dust scattering on the other hand may lead to overestimation of the mass \citealt{Zhu2019}. Some of the Herbig disks in this sample might not have formed a gas giant planet at large separations, which would explain the lower dust masses of some of the disks. This illustrates that if higher mass stars are formed with on average higher mass disks, the change in observable consequences (i.e., continuum flux and therefore inferred mass) during the evolution of the disk is likely to amplify the original trend.


\subsection{Group I and Group II}
\label{subsec:g1_vs_g2}
How do the dust mass and size correspond to other disk properties of our Herbig sample? One such property is the Meeus groups: group I and group II \citep{Meeus2001}. Different explanations for the origin of the two groups exist. Evolutionary arguments have been put forward, for example grain growth and/or settling causing group I to evolve into group II \citep{Dullemond2004a, Dullemond2004b, Dullemond2005}. However, the finding that many group I sources have disks with gaps \citep[e.g.,][]{Honda2012} caused \citet{Maaskant2013} to propose an evolutionary scenario where both groups are preceded by a primordial flaring disk. Nonetheless, \citet{Menu2015} pointed out that there were group II sources with gaps as well (based on VLTI observations, referring to gaps on au scales, as opposed to tens of au for ALMA),  giving rise to the idea that group I has an influx of evolving group II gapped disks.

Table \ref{tab:Herbig_params} contains the groups in which each Herbig disk falls following the classification of \citet{Meeus2001}. Figure \ref{fig:meeus_groups_overview} combines the dust mass and radius with the Herbig group classification and information on the structures visible in the disk. From this figure, it is apparent that the mean dust mass of the group I disks is larger than that of the group II disks. Using the same approach as before, the cumulative and fitted log-normal distributions of the group I and group II sources are shown in Fig. \ref{fig:CDF_to_PDF_meeus}. The fitting results are listed in Table \ref{tab:fit_params_meeus}. While the number of disks for each distribution is small, this analysis clearly shows a higher mean dust mass for the group I objects compared to the group II objects. Moreover, the distribution of the group II objects is similar to the distribution of the disks in Lupus (see Fig. \ref{fig:CDF_to_PDF_meeus}).

\begin{table}[b!]
\caption{Log-normal distribution fit results for the group I and group II dust mass cumulative distributions shown in Fig. \ref{fig:CDF_to_PDF_meeus}. The $M_\text{dust}$ parameters are given in log$_{10}$($M/M_\oplus$).}
\centering
\begin{tabular}{l|cc|cc}
\hline\hline
          & \multicolumn{2}{c|}{$M_\star$ ($M_\odot$)} & \multicolumn{2}{c}{$M_\text{dust}$}   \\ \hline
          & $\mu$         & $\sigma$      & $\mu$                  & $\sigma$               \\ \hline
Group I   & 2.73          & 2.99          & 1.57$^{+0.06}_{-0.07}$ & 0.34$^{+0.08}_{-0.07}$ \\
Group II  & 2.13          & 0.65          & 0.61$^{+0.15}_{-0.19}$ & 0.68$^{+0.18}_{-0.15}$ \\ \hline
\end{tabular}\\
\label{tab:fit_params_meeus}
\end{table}

As many as 13 out of 19 group I disks are resolved, and they all show structure. The resolved group II disks are smaller than the group I disks, but there are group I disks with sizes very similar to several group II disks, and in these cases both the group I and II disks show structure (for a comparison see Fig. \ref{fig:g1_vs_g2}). While there are large upper limits for both group I and group II disks, there are a few much smaller and thus more stringent dust radius upper limits for both groups, showing that there are a few much smaller disks in either group. Based on the grey region showing where most of the disks lie (as studied by \citealt{vanderMarel2021}) we predict that the five unresolved group II disks at $\sim1$~M$_\oplus$ (BF~Ori, HD~37258, HD~58647, HD~141569 and VV~Ser) are going to be 10--30~au in size when resolved. While for four of these disks the size is unknown, from visibility modeling HD~141569 is thought to have an outer disk component out to 300~au, though it is much fainter than an inner disk component with a radius of $45$~au \citep{White2018}. If these disks are indeed $\sim10$~au in size, a resolution of at least $0.02^{\prime\prime}$ is needed to resolve all five disks. This resolution is achievable with ALMA in the most extended array in Band 7.

In general, one ring is visible for most of the group I disks (9/12), while multiple rings are visible for most group II disks (3/5). Three of these resolved group II disks can be paired with group I disks of very similar size and mass (indicated by the numbered pairs 1-3 in Fig. \ref{fig:meeus_groups_overview}). Figure \ref{fig:g1_vs_g2} shows these six disks, three for each group, with similar sizes and masses for each pair. When putting these disks side-by-side, an intriguing trend appears. The group II disks are more compact, and the bulk of their mass is close to the star. This is in contrast to the group I disks, where mostly large rings are visible, with little mass close to the star. The same is found in scattered light images \citep{Garufi2017}.

As we hypothesized in \S\ref{subsec:Rdust_Mdust}, the larger disks could be connected to the formation of gas-giant planets at large separations, which can open up gaps and inward transport of material is halted. The difference between group I and group II disks could be caused by this as well: group I disks were able to form gas giants at large distances, while group II disks were not, only halting the radial drift at closer-in distances. The group I disks keep much of the dust mass at larger radii, while the inner part of the disk accretes onto the star, leaving a single large ring. This agrees well with the evolutionary scenario posed by \citet{Menu2015} and \citet{Garufi2017}, who suggested that group II disks may evolve into group I disks. The group II disks might form (giant) planets closer-in or later on, eventually creating a gap and moving towards group I. This would also explain the two other resolved group II disks not yet mentioned: AK~Sco and HD~9672, both of which show one ring and are rather old. Furthermore, \citet{Cieza2021} propose a similar evolutionary scenario based on the brightest disks from the DSHARP \citep{Andrews2018b} and ODISEA programs. \citet{Cieza2021} propose that disks first form with shallow gaps, which become more pronounced once planets start to grow; the dust is trapped at the inner edges of the rings while the inner dust disk dissipates, creating a single-ringed disk. This matches well with what we see here.

Another possibility which should be mentioned is that close (sub)stellar binary companions would have a similar effect on the disk by creating an inner hole \citep{vanderMarel2021}. All three group I sources shown in Fig. \ref{fig:g1_vs_g2} are known to be wide binaries (HD~100453 \citealt{Chen2006}; CQ~Tau \citealt{Thomas2007}; HD~135344B \citealt{Coulson1995}). While wide separation binaries are unlikely to affect disk structure (\citealt{Harris2012}, but see \citet{Rosotti2020} who suggest that the spirals in HD~100453 are caused by the companion), for HD~135344B and CQ~Tau there is indirect evidence from spirals that there could be a massive companion carving the inner gap \citep{vanderMarel2016, Wolfer2021}. A similar scenario has also been suggested for AB~Aur \citep{Poblete2020} . In HD~142527, an M-dwarf has been found to reside in the disk \citep[e.g.,][]{Lacour2016, Claudi2019}, giving rise to asymmetric structures in the disk \citep{Garg2021}. Hence, binarity at relatively close separations could also possibly explain the differences between the group I and group II sources.

The fact that we detect more massive group I disks compared to group II could then be a consequence of the spatial distribution of the dust in the disk. As Fig. \ref{fig:groups_vs_mass} shows, the distribution of stellar masses is the same in group I and group II objects and therefore, following the $M_\star$--$M_\text{dust}$ relation, they should have a similar range in dust masses. Nevertheless, the group II disks are smaller and are therefore prone to being optically thick due to most mass being in the inner regions of the disk (which is indeed what Fig. \ref{fig:g1_vs_g2} shows), reducing the {inferred} dust mass in the disk. Consequently, in disks without gaps, the stellar radiation is potentially `trapped' close to the star because much more dust is present in these close-in regions, leaving most of the disk colder and flatter, resulting in a group II-type SED. As Fig. \ref{fig:CDF_to_PDF_meeus} shows, the distribution of the disks in Lupus overlaps well with the group II objects. This overlap could originate from group II Herbig disks being unable to form massive exoplanets at large distances, like T~Tauri stars, reducing the inferred disk mass.

Lastly, in addition to the stellar masses being evenly distributed between the group I and II objects for both the observed and unobserved disks within 450~pc (see Fig. \ref{fig:groups_vs_mass}), $\sim50\%$ of all Herbig disks within 450~pc belong to group I \citep{Vioque2018} which is comparable to the $53\%$ of the Herbig disks in our work. We therefore do not expect a change in the distributions presented in Fig. \ref{fig:CDF_to_PDF_meeus} if all disks are included in the distribution.



\section{Conclusion}
\label{sec:conclusion}
We present a first look at the distribution of dust mass and the extent of disks around Herbig stars by collecting ALMA archival data for 36 Herbig disks within 450~pc. Of these, 17 disks are resolved. Two Herbig disks are undetected. Excluding the Orion star-forming region, the completeness rate is $64\%$. We compare the results to previous population studies of Lupus \citep{Ansdell2016, Ansdell2018} and Upper~Sco \citep{Barenfeld2016, Barenfeld2017}. Our results can be summarized as follows:

\begin{enumerate}
    \item The mean dust mass of the Herbig sample is a factor of about three higher than the dust masses in Lupus and a factor of about seven higher than the dust masses in Upper~Sco. The Herbig disk masses are distributed over the same range as T~Tauri disks, but are skewed towards higher masses.
    \item Herbig disks are generally larger than T~Tauri disks, although the largest Herbig disks are of similar size to the largest T~Tauri disks.
    \item The masses of the Herbig disks are as expected based on the $M_\star$--$M_\text{dust}$ relation. After scaling the dust masses following this relation to a star of 1$M_\odot$, the mean dust mass overlaps with the Upper~Sco SFR, which has a similar age to the Herbig star sample used in this work.
    \item Based on both larger and more massive disks, we speculate that these massive and large disks find their origin in an initial disk mass that increases with stellar mass, and that subsequent disk evolution enlarges the (observable) differences with disks around lower mass stars.
    \item Group I objects are found to be both more massive and larger than group II objects. Additionally, the dust mass distribution of the group II disks is similar to that of the Lupus T~Tauri disks. We argue that group II disks were unable to form massive companions or planets at (very) large radii and consequently much more dust is present at close-in regions compared to group I sources; this causes them to have masses and sizes similar to those of T Tauri disks, and they may trap stellar radiation at small radii, leaving most of the disk cold and flat, resulting in a group II-type SED.
\end{enumerate}

This work shows that the evolution of Herbig disks may hold important clues about the formation of gas-giant planets. A complete ALMA study of a homogeneous sample of Herbig disks at similar resolutions and sensitivities is much needed in order to further investigate their formation.

\begin{acknowledgements}
Astrochemistry in Leiden is supported by the Netherlands Research School for Astronomy (NOVA). This paper makes use of the following ALMA data: 2013.1.00498.S, 2015.1.00192.S, 2015.1.00889.S, 2015.1.01058.S, 2015.1.01301.S, 2015.1.01353.S, 2015.1.01600.S, 2016.1.00110.S, 2016.1.00204.S, 2016.1.00484.L, 2016.1.01164.S, 2017.1.00466.S, 2012.1.00870.S, 2017.1.01404.S, 2017.1.01545.S, 2017.1.01607.S, 2017.1.01678.S, 2018.A.00056.S, 2018.1.00814.S, 2018.1.01222.S, 2018.1.01309.S, 2019.1.00218.S, 2019.1.01813.S. ALMA is a partnership of ESO (representing its member states), NSF (USA) and NINS (Japan), together with NRC (Canada), MOST and ASIAA (Taiwan), and KASI (Republic of Korea), in cooperation with the Republic of Chile. The Joint ALMA Observatory is operated by ESO, AUI/NRAO and NAOJ. This work makes use of the following software: The Common Astronomy Software Applications (CASA) package \citep{McMullin2007}, Python version 3.9, astropy \citep{astropy2013, astropy2018}, lifelines \citep{DavidsonPilon2021}, matplotlib \citep{Hunter2007}, numpy \citep{Harris2020}, scipy \citep{2020SciPy-NMeth} and seaborn \citep{Waskom2021}. We thank the referee for their insightful comments which have improved this paper. Additionally, we would like to thank Rens Waters, Nienke van der Marel and Carsten Dominik for their comments as well. Lastly, our thanks goes to the European ARC node in the Netherlands (ALLEGRO), and in particular Aïda Ahmadi, for their help with the data calibration and imaging.
\end{acknowledgements}

\bibliographystyle{aa}
\bibliography{references.bib}

\begin{appendix}

\section{Previous disk population studies done with ALMA}
\label{app:previous_studies}
Table \ref{tab:population_studies} lists previous population studies done with ALMA and the number of Herbig stars (as listed in \citealt{Vioque2018}) present in each work.

\begin{table*}[t!]
\caption{Some of the disk population studies done with ALMA. The approximate ages and mean distances are shown together with the total number of targets and the number of these targets that are Herbigs in each survey.}
\tiny
\centering
\begin{tabular}{l c c c c l}
\hline\hline
Region            & Age (Myr) & Distance (pc) & N   & Herbigs & Ref. \\ \hline
Taurus            & 2         & 140           & 210 & 3       & 1    \\
Lupus             & 1--3      & 150           & 93  & 1       & 2    \\
$\sigma$ Orionis  & 3--5      & 385           & 92  & 0       & 3    \\
$\lambda$ Orionis & 5         & 400           & 44  & 1       & 4    \\
Upper Scorpius    & 5--11     & 145           & 106 & 0       & 5    \\
Corona Australis  & 1--3      & 154           & 41  & 2       & 6    \\
Ophiuchus         & 1         & 140           & 147 & 0       & 7, 8 \\
OMC1              & 1         & 400           & 49  & 0       & 9    \\
ONC               & 1         & 400           & 104 & 0       & 10   \\
Lynds 1641        & 1.5       & 428           & 101 & 0       & 11   \\
Chamaeleon I      & 2         & 160           & 93  & 0       & 12   \\
IC 348            & 2--3      & 310           & 136 & 0       & 13   \\
OMC 2             & 1         & 414           & 132 & 0       & 14   \\
NGC 2024          & 0.5       & 414           & 179 & 0       & 15   \\
\hline
This work         & 6.2±3.5   & 201±111       & 36  & 36      &      \\
\hline
\end{tabular}\\
\justifying{
\noindent \textbf{Notes:} An object is counted as an Herbig star if listed by \citet{Vioque2018}. The given age and distance for this work are the mean~±~standard deviation of the values given by \citet{Vioque2018}. \textit{References:} (1) \citet{Andrews2013}, (2) \citet{Ansdell2016}, (3) \citet{Ansdell2017}, (4) \citet{Ansdell2020}, (5) \citet{Barenfeld2016}, (6) \citet{Cazzoletti2019}, (7) \citet{Cieza2019}, (8) \citet{Williams2019}, (9) \citet{Eisner2016}, (10) \citet{Eisner2018}, (11) \citet{Grant2021}, (12) \citet{Pascucci2016}, (13) \citet{RuizRodriguez2018}, (14) \citet{vanTerwisga2019}, (15) \citet{vanTerwisga2020}.
}
\label{tab:population_studies}
\end{table*}

\section{Details on individual boundary objects}
\label{app:details_boundary_objects}
\subsection{HD~9672}
HD~9672 (or better known as 49~Cet) has been classified both as an old debris disk and a Herbig star. \citet{Zuckerman2012} classified HD~9672 as a debris disk belonging to the 40 Myr old Argus association. However, the disk surrounding HD~9672 has been found to be relatively gas rich compared to other debris disks \citep{Zuckerman1995, Moor2019}. Moreover, the origin of both the dust and the gas is still unknown. It could be second-generation dust and gas generated by collisions between planetesimals, or inherited from the original cloud \citep{Hughes2017}. Because of this uncertainty, we decided to leave it in the sample.

\subsection{Z~CMa, HD~58647, MWC~297, and HD~53367}
These four objects are quite massive, all belonging to the B spectral type which may be young massive objects instead of Herbig stars. Z~CMa is a binary \citep{Koresko1991} where the primary is a Herbig Be star \citep{Whitney1993, vandenAncker2004} and the secondary is a FU~Orionis variable \citep{Hartmann1989}. An asymmetric outflow is found in this system \citep{Garcia1999, Baines2006}. In HD~58647 outflows have been found as well \citep{Kurosawa2016}. MWC~297 has a large mass of $\sim14.5$~M$_\odot$ \citep{Vioque2018}. While \citet{Vioque2018} estimate its age to be 0.04 Myr, MWC~297 has been characterized as being a main sequence star \citep{Drew1997} surrounded by an evolved disk \citep{Manoj2007}. Lastly, HD~53367 is a binary star consisting of a massive ($\sim20$M$_\odot$) main sequence B0e star and a PMS secondary ($\sim4$M$_\odot$) \citep{Pogodin2006}. While classified as a Herbig star, even one of the original classified by \citet{Herbig1960}, it could also instead be a classical Be star \citep{Pogodin2006}.

\subsection{R~CrA}
R~CrA is still embedded in a dust envelope \citep{Kraus2009} and is likely in an early evolution phase \citep{Malfait1998}. It also has outflows and jets \citep{Rigliaco2019}. This case is on the boundary of being a Herbig star, because it may be in an earlier stage than Herbig stars, and therefore we keep it in the sample.

\section{ALMA archival data}
\label{app:archival_data_used}
In this work, a total of 36 ALMA archival datasets have been used. Their continuum observing frequencies and corresponding Project IDs can be found in Table \ref{tab:project_IDs}.

\begin{table*}[h!]
\caption{The Project IDs of the data used in this work with the corresponding observation frequency of the continuum data.}
\tiny
\centering
\begin{tabular}{lcccl}
\hline\hline
\makecell{Name \\ \hspace{1mm}} & \makecell{$\nu$ \\ (GHz)} & \makecell{Beam \\ ($^{\prime\prime}$)} & \makecell{RMS \\ (mJy beam$^{-1}$)} & \makecell{Project ID \\  \hspace{1mm}} \\ \hline
AB Aur      & 343.51 & 0.32 $\times$ 0.20 & 0.40 & 2015.1.00889.S \\
AK Sco      & 233.00 & 0.09 $\times$ 0.06 & 0.17 & 2016.1.00204.S \\
BF Ori      & 233.01 & 1.49 $\times$ 1.03 & 0.34 & 2019.1.01813.S \\
CQ Tau      & 232.37 & 0.07 $\times$ 0.05 & 0.05 & 2017.1.01404.S \\
HD 100453   & 234.19 & 0.03 $\times$ 0.02 & 0.02 & 2017.1.01678.S \\
HD 100546   & 233.00 & 0.02 $\times$ 0.02 & 0.02 & 2018.1.01309.S \\
HD 104237   & 219.56 & 1.31 $\times$ 0.76 & 2.20 & 2015.1.01600.S \\
HD 135344B  & 340.98 & 0.37 $\times$ 0.29 & 2.09 & 2012.1.00870.S \\
HD 139614   & 232.01 & 0.68 $\times$ 0.48 & 3.55 & 2015.1.01600.S \\
HD 141569   & 229.00 & 1.39 $\times$ 1.05 & 0.08 & 2017.1.01545.S \\
HD 142527   & 231.90 & 0.27 $\times$ 0.24 & 1.55 & 2015.1.01353.S \\
HD 142666   & 232.60 & 0.03 $\times$ 0.02 & 0.03 & 2016.1.00484.L \\
HD 163296   & 232.60 & 0.05 $\times$ 0.03 & 0.06 & 2016.1.00484.L \\
HD 169142   & 232.50 & 0.22 $\times$ 0.15 & 0.27 & 2015.1.01301.S \\
HD 176386   & 224.14 & 0.43 $\times$ 0.32 & 0.20 & 2015.1.01058.S \\
HD 245185   & 248.00 & 0.31 $\times$ 0.30 & 0.24 & 2017.1.00466.S \\
HD 290764   & 344.30 & 0.07 $\times$ 0.05 & 0.15 & 2017.1.01607.S \\
HD 31648    & 218.00 & 0.15 $\times$ 0.09 & 0.33 & 2016.1.01164.S \\
HD 34282    & 234.19 & 0.25 $\times$ 0.22 & 0.29 & 2015.1.00192.S \\
HD 36112    & 232.01 & 0.47 $\times$ 0.33 & 0.55 & 2015.1.01600.S \\
HD 37258    & 233.01 & 1.84 $\times$ 1.24 & 0.27 & 2019.1.01813.S \\
HD 53367    & 225.51 & 0.48 $\times$ 0.39 & 0.14 & 2018.1.00814.S \\
HD 58647    & 225.51 & 0.47 $\times$ 0.39 & 0.14 & 2018.1.00814.S \\
HD 9672     & 232.50 & 1.69 $\times$ 1.16 & 0.09 & 2018.1.01222.S \\
HD 97048    & 233.70 & 0.38 $\times$ 0.18 & 0.91 & 2015.1.00192.S \\
HR 5999     & 232.01 & 0.53 $\times$ 0.51 & 0.67 & 2015.1.01600.S \\
KK Oph      & 232.01 & 0.58 $\times$ 0.51 & 0.66 & 2015.1.01600.S \\
MWC 297     & 217.51 & 0.42 $\times$ 0.31 & 3.81 & 2018.1.00814.S \\
R CrA       & 231.32 & 7.48 $\times$ 4.68 & 11.3 & 2018.A.00056.S \\
TY CrA      & 224.14 & 0.43 $\times$ 0.32 & 0.20 & 2015.1.01058.S \\
V1787 Ori   & 233.02 & 1.48 $\times$ 0.99 & 0.53 & 2019.1.01813.S \\
V599 Ori    & 233.02 & 1.42 $\times$ 1.00 & 1.54 & 2019.1.01813.S \\
V718 Sco    & 232.01 & 0.87 $\times$ 0.72 & 0.92 & 2015.1.01600.S \\
V892 Tau    & 242.68 & 0.23 $\times$ 0.16 & 1.28 & 2013.1.00498.S \\
VV Ser      & 239.35 & 1.43 $\times$ 0.97 & 1.03 & 2019.1.00218.S \\
Z CMa       & 224.26 & 0.20 $\times$ 0.17 & 0.22 & 2016.1.00110.S \\

\hline
\end{tabular}
\label{tab:project_IDs}
\end{table*}

\section{Are there selection effects in our sample?}
\label{app:completeness_and_bias}

\begin{figure}[t!]
    \centering
    \includegraphics[width=0.5\textwidth]{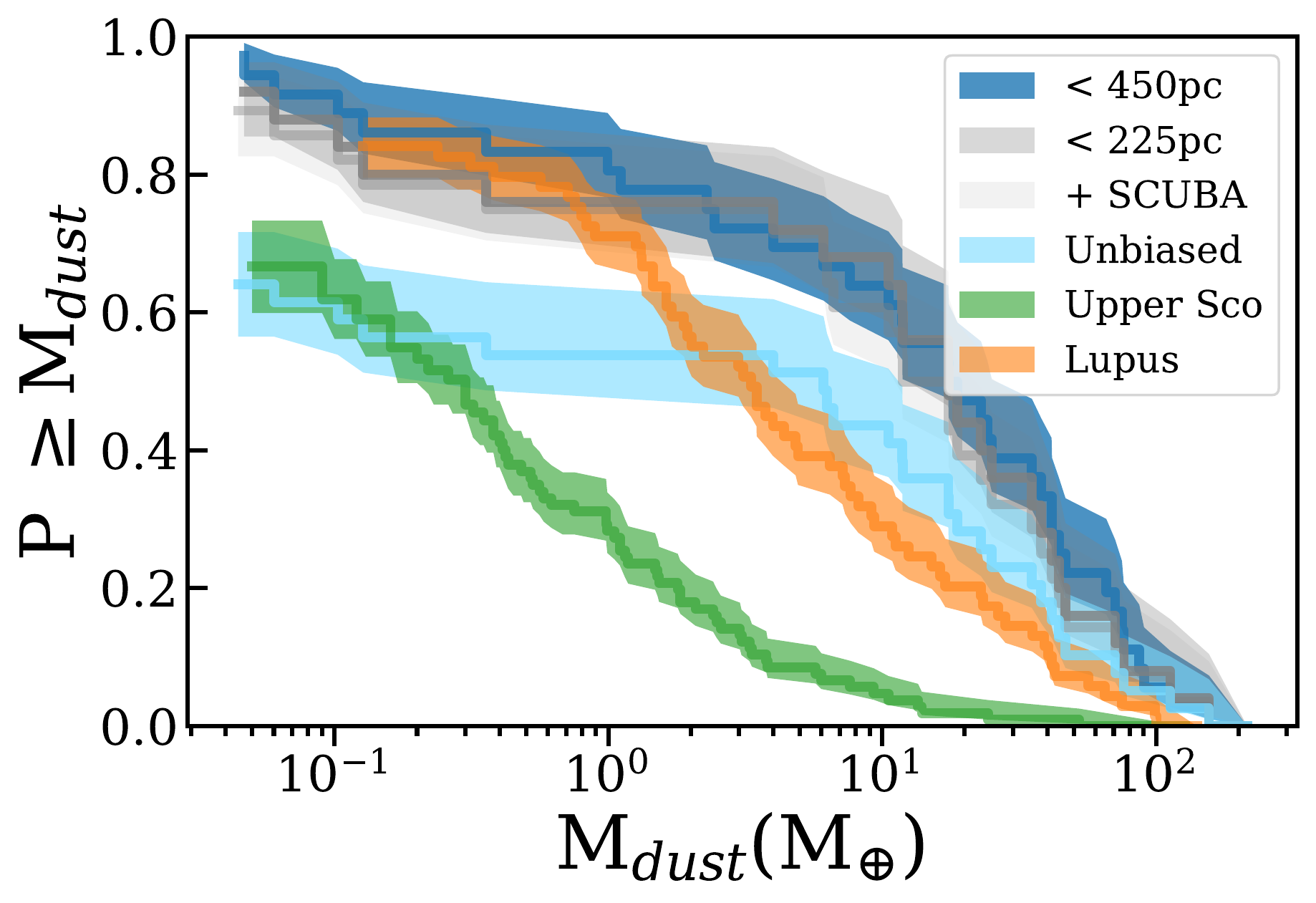}
    \caption{Cumulative distributions of the full Herbig sample ($<450$ pc), the sample within 225~pc, the sample with the additional 3 JCMT/SCUBA data within 225~pc and the unbiased version of this distribution.}
    \label{fig:CDF_SCUBA}
\end{figure}

\renewcommand{\arraystretch}{1}
\begin{table*}[t!]
\caption{Coordinates and spectral types of the Herbig stars used in this work observed with JCMT/SCUBA and the calculated flux densities and dust masses of each Herbig disk.}
\tiny
\centering
\begin{tabular}{l r c c c c c c l c c}
\hline\hline
\makecell{Name \\ \hspace{1mm}} & \makecell{R.A.$_\text{J2000}$ \\ (h:m:s)} & \makecell{Decl.$_\text{J2000}$ \\ (deg:m:s)} & \makecell{Sp.Tp. \\ \hspace{1mm}} & \makecell{Ref. \\ \hspace{1mm}} & \makecell{$F_\text{cont.}$ \\ (mJy)} & \makecell{$M_\text{dust}$ \\ ($M_\oplus$)} \\ \hline
HD 35187$^\alpha$ & 05:24:01.2 & +24:57:37 & A2 & 1 & 100.2 & $6.63\pm0.75$  \\
HD 41511$^\alpha$ & 06:04:59.1 & -16:29:04 & A3 & 2 & <0.91 & <0.044  \\
HD 150193 & 16:40:17.9 & -23:53:45 & B9.5 & 3 & 104.9  & $6.28\pm0.66$ \\
\hline
\end{tabular}\\
\textbf{Notes:} All distances are obtained from \citet{Vioque2018}. The luminosities, stellar masses, and ages were obtained from \citet{Wichittanakom2020} except the objects marked with an $\alpha$ of which these parameters were obtained from \citet{Vioque2018}. \textit{Spectral type references:} (1) \citet{Manoj2006}, (2) \citet{Jaschek1991}, (3) \citet{Levenhagen2006}.

\label{tab:Herbig_params_SCUBA}
\end{table*}

\begin{figure}[t!]
    \centering
    \includegraphics[width=0.5\textwidth]{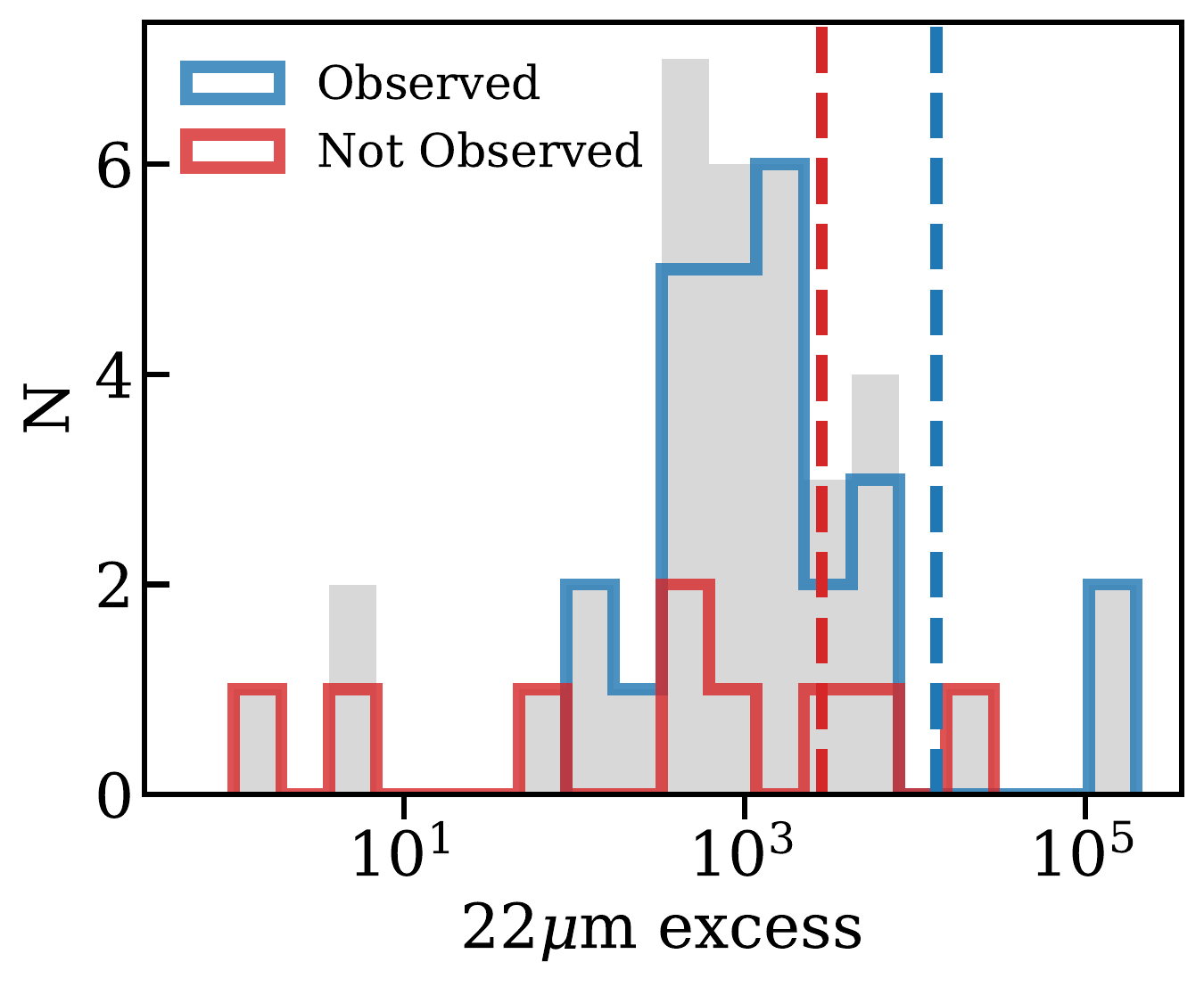}
    \caption{Distribution of WISE $22\mu$m-excess for the Herbig disks at a distance $<225$~pc. The mean $22\mu$m-excess is shown with the dashed lines.}
    \label{fig:WISE_IR_excess}
\end{figure}

Section \ref{sec:results} found that Herbigs observed with ALMA are both larger and more massive than samples of disks in Lupus and Upper~Sco. Are these Herbig disks in general indeed larger and more massive, or does the sample of Herbig disks observed with ALMA (i.e., the sample used in this work) mainly consist of more massive and larger disks?



\begin{table*}[h!]
\caption{Proposal IDs of the SCUBA data used in this work with the corresponding observation frequency of the continuum data.}
\tiny
\centering
\begin{tabular}{lclcl}
\hline\hline
\makecell{Name \\ \hspace{1mm}} & \makecell{$\nu$ \\ (GHz)} & \makecell{Instrument \\ \hspace{1mm}} & \makecell{RMS \\ (mJy beam$^{-1}$)} & \makecell{Proposal ID \\  \hspace{1mm}} \\ \hline
HD 35187      & 347.79 & SCUBA & 17.37 & m97bc36  \\
HD 41511      & 352.71 & SCUBA-2 & 0.36 & M17BL002 \\
HD 150193     & 347.38 & SCUBA & 16.23 & m01bi09  \\
\hline
\end{tabular}
\label{tab:project_IDs_SCUBA}
\end{table*}

Observations of individual objects in our sample were likely spurred by their brightness or the presence of structure in the disk either seen in scattered light or (sub)millimeter observations \citep[e.g.,][]{Fukagawa2004, Fukagawa2006, Doering2007, Tang2012, Quanz2013}. While this may have biased our ALMA sample, a sufficiently large fraction of all Herbig disks (at least out to 225 pc) have ALMA archival data, even if we consider the unobserved disks to be small and too low in mass (which future ALMA observations may determine). The sample up to 225~pc, which excludes Orion, is 64\% covered with ALMA (25/39 objects) compared to 38\% (36/96) for the 450~pc distance sample. Figure \ref{fig:CDF_SCUBA} shows that the cumulative distribution does not significantly change when only using the more complete nearby sample (i.e., all Herbig disks observed with ALMA within 225~pc) compared to the full sample (i.e., all Herbig disks observed with ALMA within 450~pc). Both the most and least massive disks are within 225~pc and hence the overall cumulative distribution does not change significantly. Therefore, the general trend that the dust masses are larger compared to those of T~Tauri stars still holds. Of the 14 objects within 225~pc that have not been observed with ALMA, 3 have been observed with JCMT/SCUBA. For these three objects, SCUBA and SCUBA-2 data products\footnote{\url{http://www.cadc-ccda.hia-iha.nrc-cnrc.gc.ca/en/jcmt/}} were used to determine the dust masses; see Table \ref{tab:Herbig_params_SCUBA}. Their continuum observing frequencies and corresponding Project IDs can be found in Table \ref{tab:project_IDs_SCUBA}. These three datasets increase the percentage of observed disks to $\sim72\%$ within 225~pc. Because of the large beam of JCMT/SCUBA ($\sim14^{\prime\prime}$), the dust radii were not determined. HD~41511 is observed with SCUBA-2 and is not detected, so only an upper limit is given. The found dust masses are low: HD~35187, HD~41511, and HD~150193 are $6.63\pm0.75$~M$_\oplus$, $<0.044$~M$_\oplus,$ and $6.28\pm0.66$~M$_\oplus$ respectively. These three extra masses do not change the overall distribution much, although they provide anecdotal evidence that unobserved disks may indeed be smaller and less massive, see Fig. \ref{fig:CDF_SCUBA}.


Assuming that all unobserved disks (so the left over 29\%) have an upper limit on the detected mass as low as the lowest upper limit in the sample, we can see what the least favorable distribution would be if the unobserved disks turned out to have low masses. This `unbiased' distribution is shown in Fig. \ref{fig:CDF_SCUBA} in light blue. It shows that the tail end of the cumulative dust mass distribution would still lie above the Lupus distribution and far above the Upper~Sco distribution which has a similar age to the Herbig disks. However, this is most likely a worst-case scenario, because in terms of 22~$\mu$m emission tracing the warm inner disk, these objects do not stand out with respect to the detected sources. The $22\mu$m excess of the disks within 225~pc with and without ALMA data are shown in Fig. \ref{fig:WISE_IR_excess}. These IR excesses were obtained by \citet{Vioque2018} from the Wide-field Infrared Survey Explorer (WISE) which is an infrared all-sky survey and therefore contains observations of most of the Herbig disks that were not observed with ALMA. A two-sided t-test using the \emph{SciPy} package \citep{2020SciPy-NMeth} shows that the null hypothesis of both distributions being the same cannot be rejected ($p$-value=0.425). While the $22\mu$m excess originates from inner parts of the disk, and no clear connection is present between this excess and the amount of dust, it does indicate that a disk is present around the star. We therefore do not expect the distribution to change significantly when including all Herbig disks within 450~pc as listed by \citet{Vioque2018}. We therefore conclude that the derived mass distribution of Herbig disks is unlikely to deviate significantly from the real one, and that the thus-far unobserved targets are likely to retain at least some disk material. Future observations will need to establish this firmly.

\end{appendix}

\end{document}